%% file: draft_v5-2.tex
\documentclass[a4paper,11pt]{article}
\pdfoutput=1 

\usepackage{jheppub} 

\usepackage[T1]{fontenc} 
\usepackage{units}
\usepackage{subcaption}
\usepackage{graphicx}
\usepackage{xcolor}
\usepackage{float}
\RequirePackage[dvipsnames]{xcolor}
\usepackage{xspace}    
\usepackage{defs}

\title{\boldmath Mapping jet substructure in heavy-ion collisions with track functions}

\author[a]{Jo\~{a}o Barata,}
\author[b]{Yeonju Go,}
\author[c,d]{Daniel Pablos}

\preprint{CERN-TH-2026-148}

\affiliation[a]{CERN, Theoretical Physics Department, CH-1211 Geneva 23, Switzerland}
\affiliation[b]{Physics Department, Brookhaven National Laboratory, Upton, NY, 11973, USA}
\affiliation[c]{Departamento de Física, Universidad de Oviedo, Avda. Federico García Lorca 18, 33007 Oviedo,
Spain
}
\affiliation[d]{Instituto Universitario de Ciencias y Tecnologías Espaciales de Asturias (ICTEA), Calle de la
Independencia 13, 33004 Oviedo, Spain}

\emailAdd{joao.lourenco.henriques.barata@cern.ch,ygo@bnl.gov,pablosdaniel@uniovi.es}

\abstract{
We investigate the dynamics of high-energy QCD cascades in heavy-ion collisions, focusing on modifications to jets' substructure induced by the presence of a quark gluon plasma (QGP). To this end, we study the properties and scale evolution of track functions, non-perturbative objects encoding the flow of energy from an initiating parton to all charged hadrons. In contrast to the more standard fragmentation functions, these objects have a non-linear renormalization group (RG) evolution, being sensitive to the entire jet fragmentation process. Using the \jewel and \hybrid Monte-Carlo models of jet quenching, we find that in both frameworks the higher moments and cumulants of the track functions' distributions exhibit sizable deviations from their vacuum baselines, demonstrating that medium-induced energy loss imprints itself in the in-medium jet fragmentation pattern. We find that the quantitative magnitude of these modifications differs significantly between the two models. This identifies track functions as a class of observables capable of discriminating between competing microscopic pictures of jet-medium interactions. 
We further examined 
the (in-medium) RG flows for the leading moments, which remain in qualitative agreement with the in-vacuum evolution. This provides an avenue to directly test the RG flows inside a QGP, linking heavy-ion jet measurements to basic QFT understanding of partonic branching.
Finally, we comment on the feasibility and challenges associated with extracting track functions from experimental data.
}

\begin{document} 
\maketitle
\flushbottom

\section{Introduction}\label{sec:intro}

Jets are among the most powerful tomographic probes of the hot and dense QCD matter created in ultrarelativistic heavy-ion collisions~\cite{Busza:2018rrf,Mehtar-Tani:2013pia}. Produced at the earliest moments of the collision, jets subsequently fragment while traversing the rapidly expanding plasma, thereby encoding information about the spacetime evolution of the bulk medium~\cite{Apolinario:2022vzg}. Although realizing this tomographic program remains challenging, few other observables offer comparable sensitivity to the dynamical properties of the quark gluon plasma (QGP), making jets a unique probe of the bulk. Developing a theoretically consistent and experimentally testable picture of in-medium jet fragmentation is therefore essential for advancing our understanding of QCD under extreme conditions.

Despite significant recent progress~\cite{Mehtar-Tani:2025rty,Cao:2020wlm,Cao:2024pxc}, the current theoretical description of jet fragmentation in heavy-ion collisions remains incomplete, with many aspects still guided primarily by phenomenological considerations. Efforts that strengthen the connection between first-principles QCD dynamics and experimentally accessible jet observables are thus highly desirable. In recent years, sustained progress has been made toward a more comprehensive theoretical description with developments including: probing the medium-modified virtuality cascade~\cite{Caucal:2018dla, Cunqueiro:2023vxl, Mehtar-Tani:2017ypq,Mehtar-Tani:2017web,Kumar:2025rsa}, describing jet evolution in structured and time-dependent backgrounds~\cite{Barata:2023qds,Kuzmin:2023hko,Boguslavski:2024ezg, Carrington:2021dvw, Adhya:2020xcb, Silva:2025dan,Barata:2024bqp,Hauksson:2023tze,Carrio:2026xrr}, and determining the pattern of \textit{bremsstrahlung} radiation induced by the medium~\cite{Arnold:2020uzm,Arnold:2025dqj,Abreu:2024wka, Barata:2024xwy, Casalderrey-Solana:2012evi, Mehtar-Tani:2012mfa, Blaizot:2013vha,Kuzmin:2025fyu}, among other advances. Collectively, these developments aim to provide a robust theoretical framework capable of confronting the increasingly more differential experimental data becoming available, see e.g.~\cite{CMS:zgPbPb:2018,ALICE:thetaGzgPbPb:2022,ATLAS:rgSuppressionPbPb:2023,
CMS:gammaJetRgGirthPbPb:2025,ATLAS:largeRSubstructurePbPb:2023,
CMS:groomedMassPbPb:2018,ALICE:massAngularitiesPbPb:2025,ALICE:2026giw,ALICE:2026zyx}.

\begin{figure}[h!]
    \centering
    \includegraphics[width=0.65\linewidth]{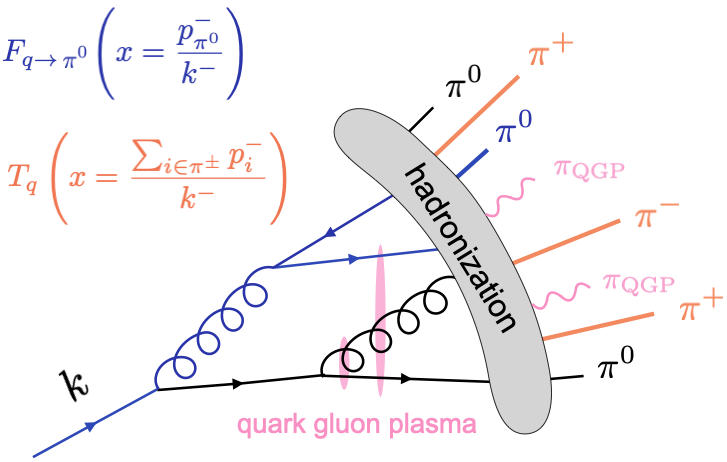}
    \caption{Illustration depicting the fragmentation of an initial high-energy quark with momentum $k$ to pions, with $k^-$ indicating its (light-cone) energy. For the fragmentation function $F_{q\to \pi^0}$, we indicate in blue the final state hadron carrying energy fraction $x$. The partons entering the evolution of the fragmentation function are also indicated in blue, making explicit that $F_{q\to \pi^0}$ is not sensitive to the full fragmentation process. Contrastingly, in orange we indicate the final state charged pions contributing to the track function $T_q$. As the track function measures the energy flow to all charged hadrons, it is sensitive to the entire fragmentation pattern. We further indicate the interactions with the QGP via pink vertical blobs. Finally, we also take into account pions which emerge from the bulk due to the jet propagation; these would correspond to medium response contributions, which we discuss in the Appendix. }
    \label{fig:cartoon}
\end{figure}

In this work, we aim to further expand on these recent developments by motivating a new tool --- the track functions $T_i(x)$ --- to resolve the \textit{hard} jet substructure, allowing to understand how inter-parton correlations build up during the branching process in the medium. At a phenomenological level, track functions can be understood as measuring the energy fraction $x$ carried by final state charged hadrons,\footnote{In fact, other conserved quantum number can be chosen instead. The choice of electric charge motivates the naming of these non-perturbative functions.} coming from an initiating parton $i$, see Fig.~\ref{fig:cartoon}. As a result, they are sensitive to the entire branching history inside the jet and thus constitute optimal probes of the fragmentation pattern. It is worth clarifying that the track functions, although related, are not equivalent to the more standard inclusive fragmentation functions, already discussed in the context of heavy-ion collisions~\cite{Casalderrey-Solana:2018wrw,Caucal:2020xad,Chen:2020tbl,JETSCAPE:2023hqn}. The difference between these two non-perturbative objects lies in the fact that fragmentation functions only measure the energy flowing to a finite set of hadrons, while the track functions capture the energy flow to all charged hadrons. In a perturbative setting, this is reflected directly in the renormalization group (RG) evolution which is highly non-linear for the track functions, while (single-hadron) fragmentation functions satisfy Dokshitzer--Gribov--Lipatov--Altarelli--Parisi (DGLAP) evolution~\cite{Dokshitzer:1977sg,Altarelli:1977zs}.

The leading structure and evolution of track functions was first discussed more than a decade ago~\cite{Chang:2013rca,Chang:2013iba}, as a way to match charged track's based measurements (in proton+proton collisions) with theory predictions. Their properties have since been widely explored and generalized~\cite{Li:2021zcf,Jaarsma:2022kdd,Chen:2022pdu,Chen:2022muj}, with special focus being given in the context of energy correlators~\cite{Moult:2025nhu} and generalized detectors~\cite{Jaarsma:2023ell,Chen:2020vvp,Lee:2023tkr,Barata:2025uxp,Lee:2023npz}, where track functions enter as matching coefficients to hadronic observables. Initial steps towards understanding the evolution of these objects in the context of heavy-ion collisions have also been recently taken~\cite{Barata:2024bmx,Barata:2024nqo}, where they have been shown to share common features with the partonic radiative energy loss, see e.g.~\cite{Mehtar-Tani:2024mvl,Mehtar-Tani:2017web}. Building on these recent theoretical developments, this work demonstrates that track functions provide a new window into the properties of the virtuality cascade inside jets. In particular, we argue that they offer a systematic framework for studying jet fragmentation, one that can be addressed from both theoretical and experimental perspectives, which has so far not been fully realized with other observables in heavy-ion environments.

This work is organized as follows. We start by reviewing the basic properties of track functions and their RG evolution in perturbative QCD in Section~\ref{sec:tracks_eloss}. In the following Section~\ref{sec:tracks_MC}, we present results for Monte-Carlo (MC) based simulations in heavy-ion collisions. 
We summarize our findings in Section~\ref{sec:conclusion}, also discussing their potential experimental extraction. Additional results, including the effects of medium response, are presented in the Appendix.

\section{Properties of track functions in QCD}
\label{sec:tracks_eloss}

Track functions are non-perturbative objects which measure the energy fraction $x$
carried by charged hadrons originating from an initial parent parton, see Fig.~\ref{fig:cartoon}.
They were first introduced as a theoretical framework to mimic track-based
experimental measurements~\cite{Chang:2013iba,Chang:2013rca}, and were soon applied to phenomenological studies of jet
observables at the LHC. Nonetheless, since computing the full hadronic cross-section
requires convolutions with the entire track-function distribution, their practical
application was disfavored for some time. However, with the development of energy-flow
based measurements~\cite{Moult:2025nhu}, interest in track-function–like objects has been reignited.
The reason for this is that, unlike in the case of jet shapes, for energy-flow observables the moments
of track functions act as multiplicative matching coefficients between hadronic and
partonic observables~\cite{Chen:2020vvp}, greatly simplifying their phenomenological application.
In what follows, we briefly summarize the basic properties of track functions and highlight aspects of their
renormalization group (RG) structure which will be relevant for the discussion in
later sections.

Track functions admit an operatorial definition closely related to, but generalizing,
the notion of fragmentation functions. For the case of quarks this is given by
\begin{align}
T_q(x) &= \int dy^+ d^2\mathbf{y} \,
e^{i k^- y^+/2}
\sum_{N,C}
\delta\!\left(x-\frac{P_C^-}{k^-}\right)
\frac{1}{2N_c}
{\rm Tr}\!\left[
\frac{\gamma^-}{2}
\langle \Omega|\psi(y)|N,C\rangle
\langle N,C|\bar\psi(0)|\Omega\rangle
\right] ,
\label{eq:def_tracks}
\end{align}
where $|\Omega\rangle$ is the QCD vacuum, $|N,C\rangle$ denotes the decomposition of the final state into neutral and
charged components, $P_C^-\equiv \sum_{i\in C} p_i^-$ is the total minus light-cone momentum carried by charged states, and the measurement function restricts the sum to the energy
carried by charged particles only, see Fig.~\ref{fig:cartoon}.
A similar formula can be written for gluons, and we have omitted the gauge links
required for gauge invariance. If the sum over states is extended over all final states, energy conservation implies
$T_q(x)=\delta(1-x)$. Conversely, if the measurement isolates the energy fraction
carried by a single hadron $h$, the track function reduces to the standard
fragmentation function $F_{q\to h}(x)$.
From the above definition it follows that
\begin{align}
\int_0^1 dx \, T_i(x,\mu) = 1 ,
\end{align}
which reflects the inclusive sum over a complete set of hadronic final states. Moments of the track functions are defined as
\begin{align}
T_i[N] = \int_0^1 dx \, x^N T_i(x) ,
\label{eq:T_moment}
\end{align}
following from above that $T_i[0]=1$.
The first moment measures the mean charged energy fraction, while higher moments
probe increasingly detailed features of the charged energy distribution associated to the fragmentation process from the initiating parton. As a result, since the track functions measure the entire energy flow into charged hadrons, their moments give direct access to the evolution of the particle correlations inside the jet cascade.

At sufficiently high energies the RG evolution of track functions can be computed
perturbatively~\cite{Chang:2013rca,Li:2021zcf}. At leading order, the evolution equation reads
\begin{align}
\mu\frac{d}{d\mu}T_i(x)
=
\frac{\alpha_s}{2\pi}
\sum_{j,k}
\int_0^1 dx_1
\int_0^1 dx_2
\int_0^1 dz
\,
T_j(x_1)T_k(x_2)
P_{i\to jk}(z)
\,
\delta[x-(zx_1+(1-z)x_2)] \, .
\label{eq:RG_D}
\end{align}
Here $P_{i\to jk}$ are the leading order QCD splitting kernels. Combining Eq.~\eqref{eq:T_moment} with Eq.~\eqref{eq:RG_D} one obtains the RG
evolution of the moments
\begin{align}
\mu \frac{d}{d\mu}T_i[N]
=
\frac{\alpha_s}{2\pi}
\sum_{j,k}
\int_0^1 dz
\,
P_{i\to jk}(z)
\sum_{m=0}^N
\binom{N}{m}
z^m(1-z)^{N-m}
T_j[m]T_k[N-m] \, .
\label{eq:RG_moments}
\end{align}
From a physics perspective Eq.~(\ref{eq:RG_moments}) exposes a key difference between
track functions and standard fragmentation functions: the evolution of a given moment
is not independent but mixes with all lower moments. This reflects the fact that the
measured charged energy fraction results from the entire branching history of the
cascade. Emissions at different stages of the shower therefore contribute collectively
to the observable. Fragmentation function moments, by contrast, evolve linearly and
independently, effectively probing the inclusive production of a single hadron. Track
functions thus encode genuinely collective information about the jet cascade and
provide a natural framework to describe how energy is redistributed through the
branching process.

Since the RG evolution mixes different moments of the distribution, it is convenient
to reorganize the description in terms of cumulants rather than raw moments.  The first few cumulants
can be written in terms of the moments as
\begin{align}
\label{eq:cumdef}
\kappa_1 &= T[1] \, , \nn
\kappa_2 &= T[2] - T[1]^2 , \nn
\kappa_3 &= T[3] - 3T[2]T[1] + 2T[1]^3 \, , \nn
\kappa_4 &= T[4] - 4T[3]T[1] - 3T[2]^2 + 12T[2]T[1]^2 - 6T[1]^4 \, ,
\end{align}
for either flavor. In this basis the hierarchy of the evolution equations becomes more transparent, and can be further constrained using energy conservation, uniquely fixing the form of the lowest cumulants' RG flow structure, see~\cite{Jaarsma:2022kdd} for a detailed discussion.
A useful quantity in this context is the difference between the mean charged fractions
for quarks and gluons:
\begin{align}
\Delta \equiv T_q[1] - T_g[1] \, .
\end{align}
Phenomenologically, since hadronization produces charged and
neutral pions in approximately isospin symmetric proportions, one has that  $T_q[1] \simeq T_g[1] \simeq 2/3$, implying $|\Delta|\ll1$.
As a result, the structure of the RG evolution for the first cumulants can be reduced to:
\begin{align}
\mu \frac{d}{d\mu}\,\kappa_2
&= \gamma_2\,\kappa_2 + \mathcal{O}(\Delta^2 ),
\nn
\mu \frac{d}{d\mu}\,\kappa_4
&= \gamma_4\,\kappa_4
+ \beta\,\kappa_2^{\,2}
+ \mathcal{O}(\Delta) \, ,
\label{eq:kappa2RG}
\end{align}
where $\gamma_i$ are the relevant anomalous scaling dimensions and $\beta$ a mixing coefficient. Equation~(\ref{eq:kappa2RG}) shows that the variance evolves approximately
multiplicatively under the RG flow. In contrast, the fourth cumulant receives a
source term proportional to $\kappa_2^2$. This term represents the first genuinely
nonlinear mixing in the cumulant hierarchy and implies that non-Gaussian features
of the charged-energy distribution can be dynamically generated even if the
initial distribution were approximately Gaussian. Importantly, no additional nonlinear structures appear at this order. Terms such as $\kappa_2\kappa_3$ or $\kappa_2^3$ carry higher cumulant order and thus first
arise in the evolution of higher cumulants. As a result, the evolution in the $(\kappa_2,\kappa_4)$ space gives direct information about the non-linear RG evolution, informing about the structure of the partonic correlations inside the cascade process. Studying the branching evolution in this plane in a QGP background thus provides direct information about modifications to the hard fragmentation process inside jets, making it possible to gauge the differences of the in-medium virtuality cascade compared to that in vacuum.

\section{Track functions from jet quenching Monte-Carlo models}\label{sec:tracks_MC}
The study of track functions in the context of heavy-ion collisions is complex to carry out using purely analytic methods, see e.g. discussion in~\cite{Barata:2024nqo}. Nonetheless, these objects can be directly studied using dedicated Monte-Carlo simulations, which include a more complete picture of the entire modified jet cascade. In what follows, we briefly introduce two jet quenching MC frameworks: the \hybrid strong weak coupling model and the \jewel 2.3 event generator. In our analysis, we identify \textit{universal} features shared by the observables in both models, while we also discuss visible discrepancies which can be potentially used for model discrimination via confrontation with future experimental data.

\subsection{Monte-Carlo selection}
The \hybrid strong/weak coupling model~\cite{Casalderrey-Solana:2014bpa,Casalderrey-Solana:2015vaa,Casalderrey-Solana:2016jvj} is based on the combination of the perturbative and non-perturbative aspects of jet evolution within a strongly coupled medium under a single framework. The high-$Q^2$ splittings that occur due to the DGLAP evolution taking place after the hard scattering are modeled using \pythia8~\cite{Bierlich:2022pfr}. A spacetime structure is ascribed to the jet shower, where each virtual parton takes a time $\tau_f=2E/Q^2$ to split~\cite{Casalderrey-Solana:2011fza}, the formation time of the dipole. In between splittings, partons can interact with the medium, as long as the local temperature is above $T>T_c$, the pseudocritical temperature of the QGP. This interaction, which is assumed to be predominantly non-perturbative as the temperature is of the order of the QCD confinement scale, $T\sim \Lambda_{QCD}$, is modeled using results obtained in the gauge/gravity duality for a non-abelian superconformal theory in the large $N_c$ limit at finite temperature~\cite{Chesler:2014jva,Chesler:2015nqz}. Hadronization of those partons that did not lose all of its energy is done using the Lund string model in \pythia8. In order to preserve energy-momentum conservation, one needs to account for the energy and momentum lost by the jet, which has been deposited in the medium. At strong coupling, this medium response takes the form of hydrodynamic wakes already at distances of the order of $1/T$~\cite{Chesler:2007sv}. These wakes particlize at the freezeout hypersurface using the Cooper-Frye procedure, yielding a cloud of soft hadrons at relatively large angles with respect to the source orientation. The main parameter of the model, the strength of the coupling that enters the strongly coupled energy loss rate, $\kappa_{sc}$, was determined by a global fit to jet and hadron data from RHIC and LHC available in 2018~\cite{Casalderrey-Solana:2018wrw}. The version of the \hybrid model using in the present paper does not include perturbative elastic scatterings~\cite{Hulcher:2026dht} nor finite resolution effects~\cite{Hulcher:2017cpt}. Their impact on the observables studied here can be assessed in future work.

In contrast with the \hybrid model, \jewel is a perturbative QCD based MC event generator for jet evolution in heavy-ion collisions that describes the interplay of shower radiation and in medium re-scattering in a microscopic framework. It takes hard scattered partons generated with \pythia6 and then simulates in medium interactions that lead to parton energy loss. The main mechanism for energy loss is bremsstrahlung, triggered by elastic collisions with the medium constituents. These interactions modify the kinematics of the jet partons and induce emissions provided that their formation time is smaller than ongoing vacuum-like emissions, emulating the Landau-Pomeranchuk-Migdal effect in QCD~\cite{Migdal:1956tc,Landau:1953gr,Landau:1953um}. This means that in \jewel even the earliest stages of the shower can be modified by medium effects, although the probability is small due to the high-momentum exchange needed (which is rare, since collisions are dominated by the infrared regulator, the Debye mass) to alter a high-virtuality, small formation time vacuum-like emission. The impact on substructure observables by recoil particles arising from the $2\to 2$ elastic scatterings, which represent medium response effects in \jewel, has been studied in~\cite{Milhano:2017nzm}. Recently, color coherence effects have been included in the model~\cite{Zapp:2026cqf}, although their repercussion to the angular distribution of recoil particles~\cite{Pablos:2024muu} has not yet been studied. In the present work we will not consider the newly introduced color coherence effects in \jewel. The \jewel version $2.3.0$ was used and the initial temperature of the medium in central collisions (impact parameter $b=0$, parameter ``TI'') is set to 0.59, and the initial proper time (parameter ``TAUI'') is set to 0.4. We set the remaining model parameters to their default values.

In both models we use the same configuration, as follows. Prompt photon+jet processes are generated in proton+proton collisions (\pp) and lead+lead collisions (\pbpb) at center of mass energy $\sqrt{s_{_{\mathrm{NN}}}}=5.02$~TeV, using 0--5\% central \pbpb events (boson-jet observables were first studied in \hybrid in~\cite{Casalderrey-Solana:2015vaa} and in \jewel in~\cite{KunnawalkamElayavalli:2016ttl}). The track function is studied as a function of the initial hard scale, denoted $\mu$ (parameter $\widehat{p}_{\mathrm{T,min}}$ in \pythia). We use samples generated at six discrete values of $\mu$ ($\mu=100,200,300,400,500,1000$~GeV) to study the scale dependence of the track functions and of their moments and cumulants. Initial-state radiation (ISR) in both \pp and \pbpb collisions is turned off in both models, and medium response contributions are disabled by turning off ``recoil'' in \jewel and ``wake'' in \hybrid. Thus, the extracted track functions primarily reflect modifications of the shower rather than contributions from the ISR and medium response. We refer an interested reader to the material in Appendix for a study of these effects on the extraction of track functions.

Jets are reconstructed at hadron level with the anti-$k_{\rm T}$ algorithm~\cite{Cacciari:2008gp} using the \fastjet package~\cite{Cacciari:2011ma}. The default jet radius is $R=0.2$ (with additional checks for $R=1.0$ reported in Appendix), where $R \equiv \sqrt{(\Delta\eta)^2+(\Delta\phi)^2}$. Jets are required to have $\ptjet>50$~GeV and to lie within $|\etajet|<2$. Only the leading jet in each event is considered, and quark-initiated and gluon-initiated jets are examined separately~\footnote{To determine whether a jet is quark- or gluon-initiated, we carry a standard leading order angular matching with the partons produced at the hard scattering. Since we only considers prompt photons, one of the partons is always a photon, and so the identity of the other parton automatically defines the flavor of our leading jet.}. For each jet, we define the charged transverse-momentum fraction $x$ as
\begin{equation}
x \equiv \frac{\left(\sum p_{\rm charged}\right)_{\rm T}}{\ptjet}\,,
\end{equation}
where the sum denotes the four-vector sum of all charged final-state particles within the jet radius, and the subscript ${\rm T}$ denotes the transverse momentum of the resulting summed four-vector. We note that this definition matches the one used in the recent measurement in \pp collisions performed by the ATLAS collaboration~\cite{ATLAS:2025qtv}. The track function is then defined as the normalized distribution of $x$ at scale $\mu$,
\begin{equation}
T(x,\mu) \equiv \frac{1}{N^{\rm jet}(\mu)}\frac{dN^{\rm jet}(\mu)}{dx}\, ,
\end{equation}
where $N^{\mathrm{jet}}$ is the number of jets.

\subsection{Extraction of track functions and track moments}

\begin{figure}[t!]
\centering    
    \includegraphics[width=0.48\textwidth]{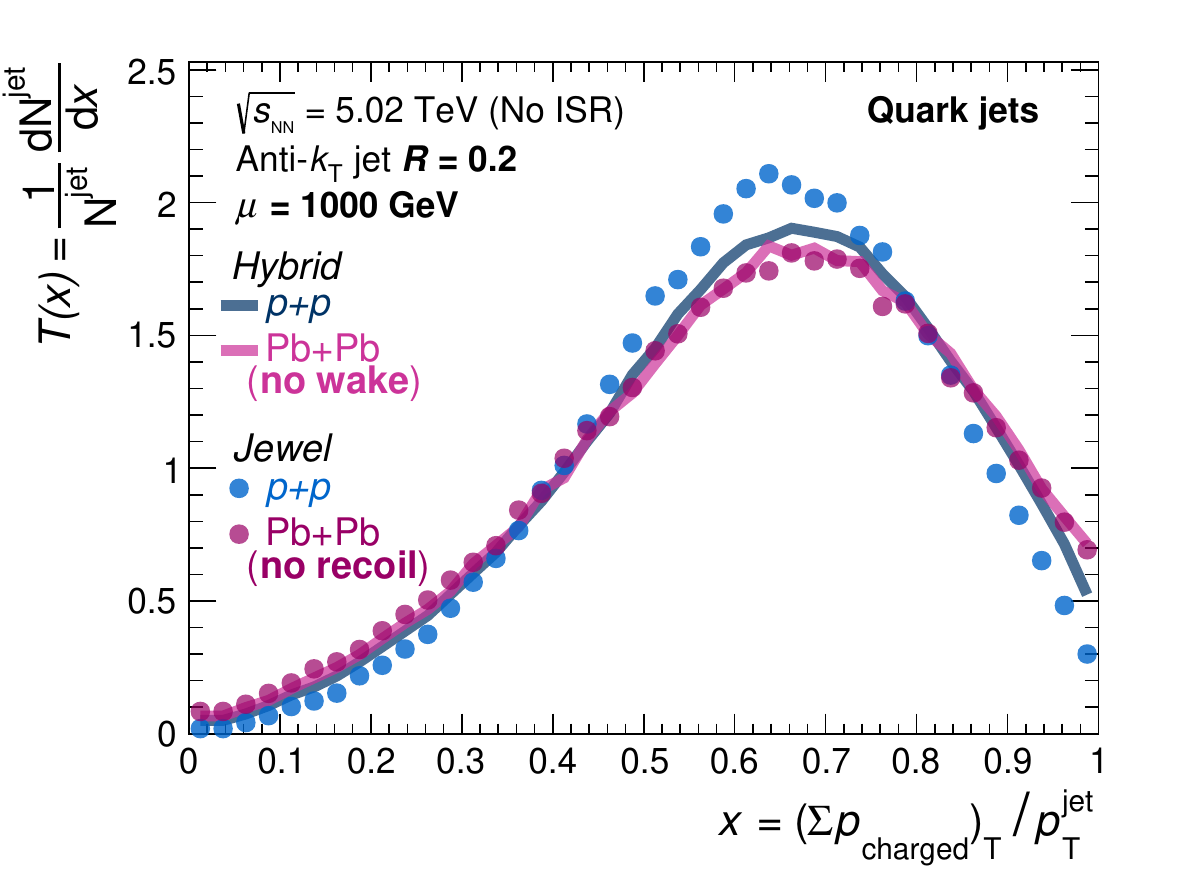}
    \includegraphics[width=0.48\textwidth]{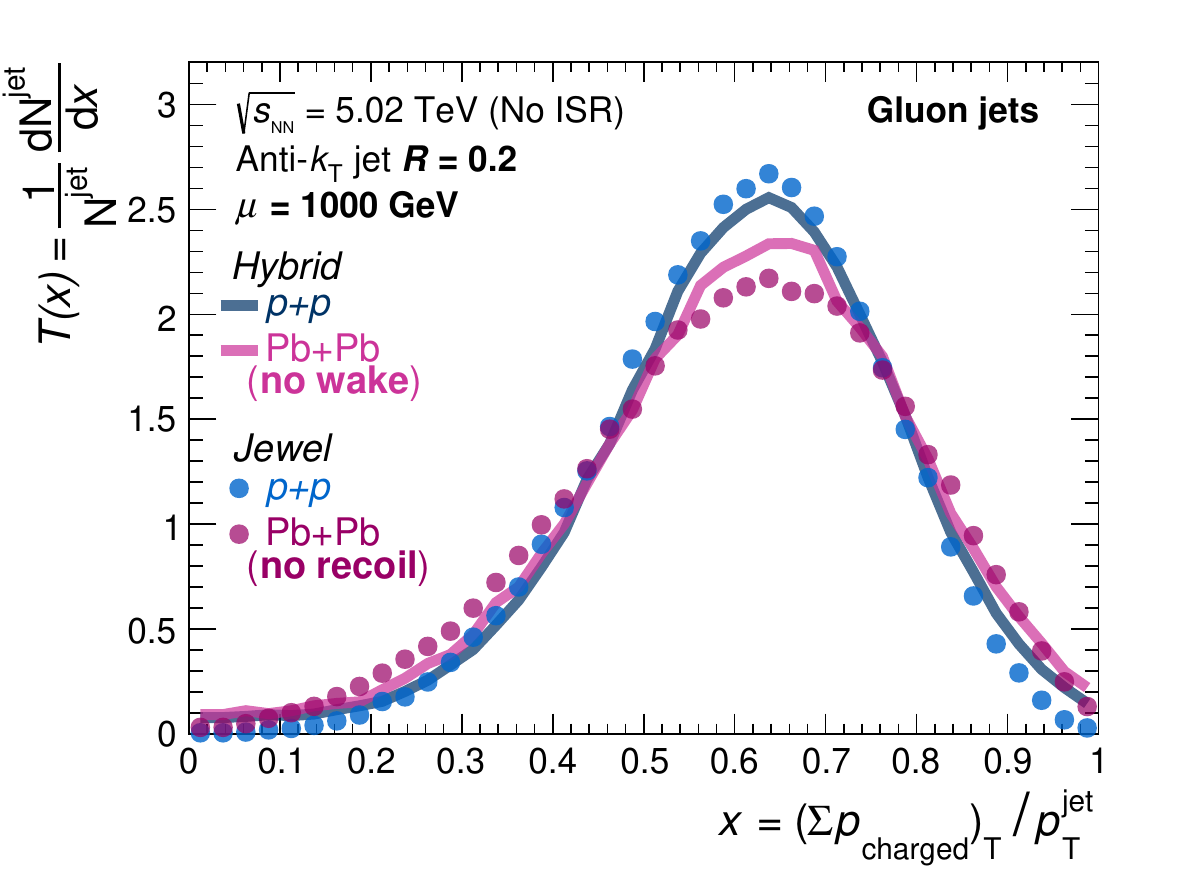}
    \caption{Track function for quark-initiated (left) and gluon-initiated jets (right) for \hybrid (lines) and \jewel (markers) in \pp and \pbpb collisions without medium response. Jets with $R$=0.2 and the $\mu$ scale is 1000 GeV.}
    \label{fig:jetTF_hybrid_jewel_pthat1000_R2}
\end{figure}

Figure~\ref{fig:jetTF_hybrid_jewel_pthat1000_R2} shows the track functions at $\mu=1000$~GeV for jets with $R=0.2$, shown separately for quark-initiated and gluon-initiated jets, in both \pp and \pbpb collisions. In all cases the distributions peak at $x\simeq 0.66$, reflecting the approximate isospin symmetry of pion production. In addition to this, for both models we observe that the quark-initiated jets have a broader distribution than gluon-initiated jets. This feature is well expected from \textit{naive} Casimir scaling arguments, which predicted gluon-initiated jets to be more highly populated, and thus be less sensitive to fluctuations. We note that even for \pp collisions, the MC models show discrepancies with respect to one another, which may arise from differences between \pythia 8 and \pythia 6. As a result, in the following analysis the comparison between models has to be carried out at a qualitative level.

For both \hybrid and \jewel, the track functions in \pbpb collisions are broader than in \pp collisions, indicating that in-medium shower modifications change the event-by-event charged fraction even when medium response is disabled. As described in Appendix, including medium response has a small effect for $R=0.2$, consistent with the expectation that the medium response component contributes predominantly at larger angles and therefore has reduced overlap with a narrow jet cone. Despite the aforementioned differences in vacuum baselines between \hybrid and \jewel, it is clear that medium-induced modifications are larger in the latter model for this observable, possibly related to the role played by semi-hard medium-induced radiation in the evolution of the parton shower.

Despite the observed broadening, the \pp and \pbpb track functions remain qualitatively similar at large evolution scales, such as $\mu=1000~\mathrm{GeV}$. This behavior is expected, since medium-induced modifications are expected to be suppressed at large momentum scales. Indeed, this corresponds to the limit where the in-medium jet \textit{resembles} a in-vacuum jet (up to a energy rescaling due to quenching), indicating that at large momentum scales one recovers an unmodified partonic cascade, see e.g.~\cite{Cunqueiro:2023vxl,CMS:2026bmz} for recent related study. The lower scale results presented in the Appendix provide a complementary illustration of this behavior, showing more pronounced deviations of the \pbpb track functions from their \pp baselines when medium effects are expected to be larger. Since track functions are sensitive to the entire cascade, the observed convergence of the \pp and \pbpb distributions at large $\mu$ give yet another clear evidence for recovery of a vacuum-like jet in heavy-ion collisions, see also~\cite{Cunqueiro:2023vxl} for recent related discussion.

\begin{figure}[t!]
\centering    
    \includegraphics[width=0.48\textwidth]{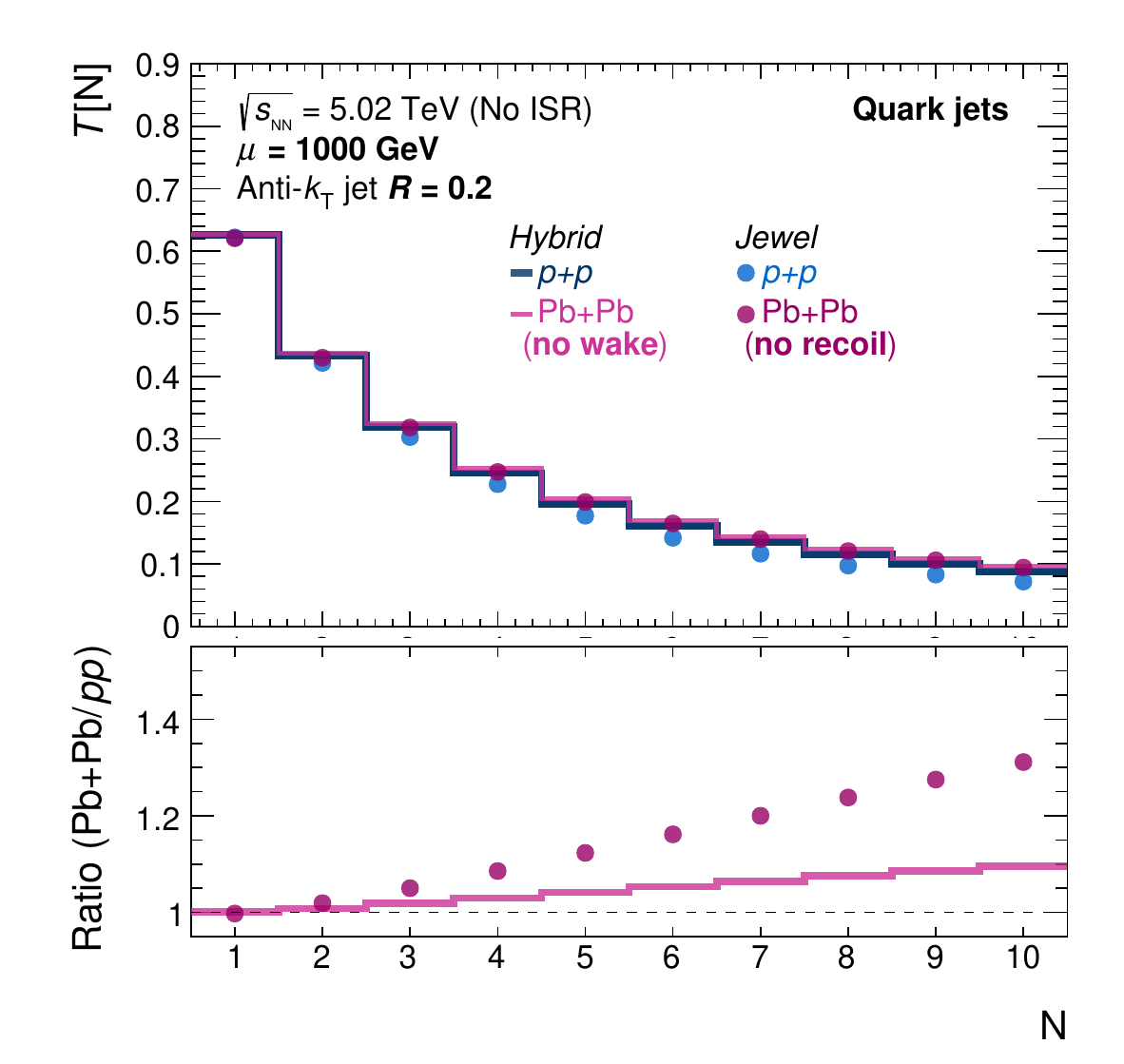}
    \includegraphics[width=0.48\textwidth]{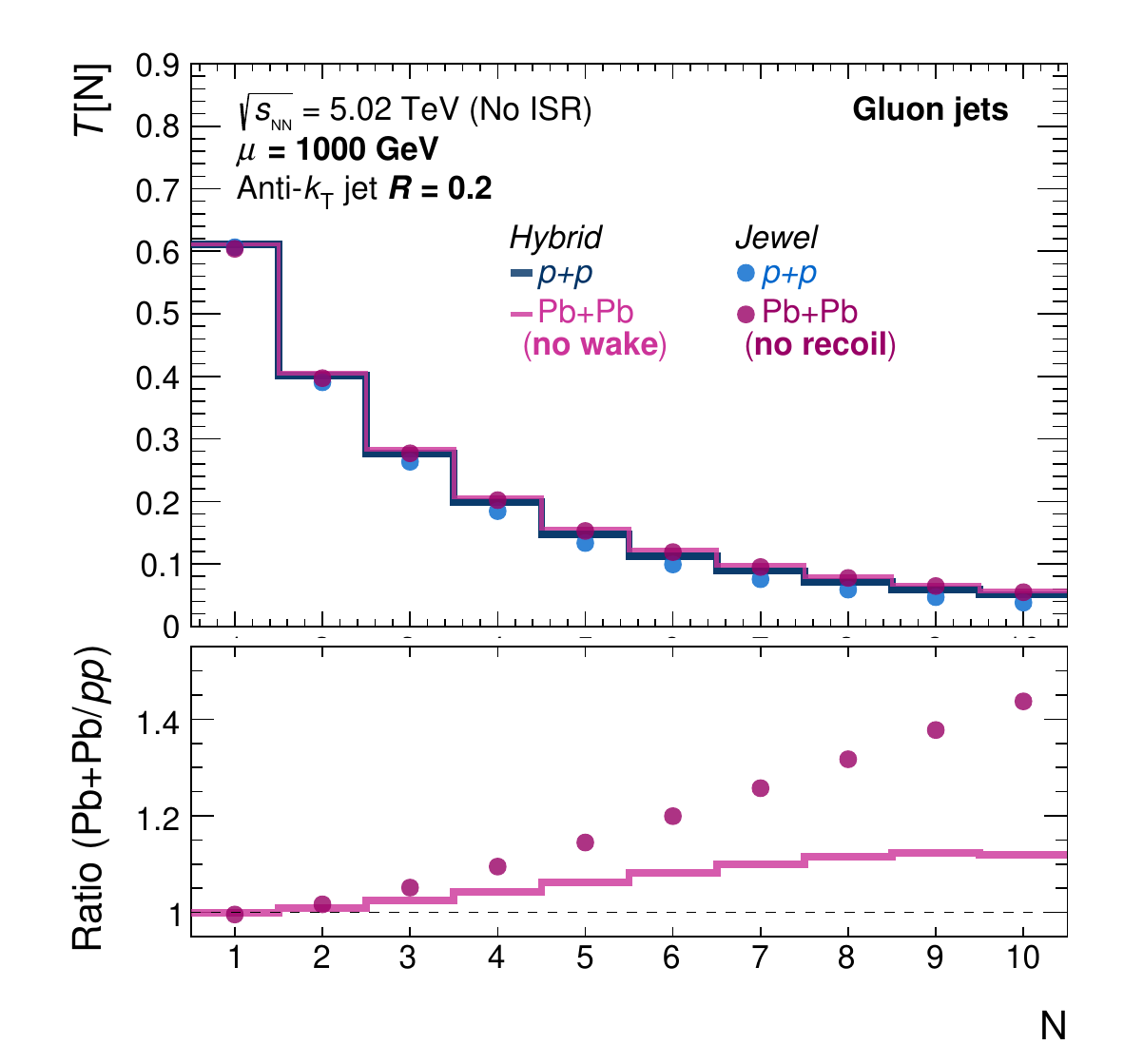}
    \caption{ The $N$-th order moments of the track function for quark-initiated (left) and gluon-initiated (right) jets with $R$=0.2. Results are shown for \hybrid (lines) and \jewel (circles) in \pp (blue) and \pbpb collisions without the medium response (pink) with $\mu$ scale of 1000 GeV. The bottom panels represent the ratio of \pbpb to \pp.}
    \label{fig:jetTF_moment_vs_N_mu1000_R2}
\end{figure}

As mentioned in Section~\ref{sec:tracks_eloss}, the study of the track function properties can be made more transparent by studying their moments and cumulants. Following Eq.~\eqref{eq:T_moment}, the moments of the track functions are directly obtained from the distributions in Fig.~\ref{fig:jetTF_hybrid_jewel_pthat1000_R2},
for $N=1$ up to $N=10$. Figure~\ref{fig:jetTF_moment_vs_N_mu1000_R2} shows $T_i[N]$ for $\mu=1000$~GeV and $R=0.2$ for quark-initiated and gluon-initiated jets, together with the \pbpb to \pp ratios. As expected, $T_i[N]$ decreases with increasing $N$ since $x\in[0,1]$. 

The \pbpb to \pp ratio increases with $N$ in both \hybrid and \jewel models, indicating that the track function in \pbpb collisions is relatively enhanced at larger charged particle fractions compared to \pp, which is emphasized by higher moments. The size of the modification differs between the generators. In \hybrid, the \pbpb to \pp ratios at $N=10$ reach only about 10\% for quark-initiated jets and about 25\% for gluon-initiated jets. In contrast, \jewel shows a substantially larger increase with $N$, with the ratio reaching about 30\% for quark-initiated jets and about 45\% for gluon-initiated jets at $N=10$. This indicates that, for the same no-medium-response configuration, \jewel predicts a stronger medium-induced reshaping of the charged fraction distribution, in consistency with the observations made from Fig.~\ref{fig:jetTF_hybrid_jewel_pthat1000_R2}, which is, as expected, even larger for gluon-initiated jets. 

The agreement of the lowest moments is expected as we define the energy fraction $x$ in \pbpb collisions with respect to the quenched jet, and thus the overall energy loss of the jet is factored out to a large extent. In the limit of a fully coherent energy loss model (this is, if the medium cannot resolve jet substructure fluctuations), this would imply that the leading momentum would agree with the vacuum sample. As a result, the observed trend clearly indicates that the presence of the medium, inducing substructure-dependent energy loss, leads to larger fluctuations in the cascade structure. In other words, the rising \pbpb to \pp ratio with increasing $N$ provides a clear indication that track function moments are sensitive to medium-induced modifications of jet structure.

\begin{figure}[t!]
\centering    
    \includegraphics[width=0.48\textwidth]{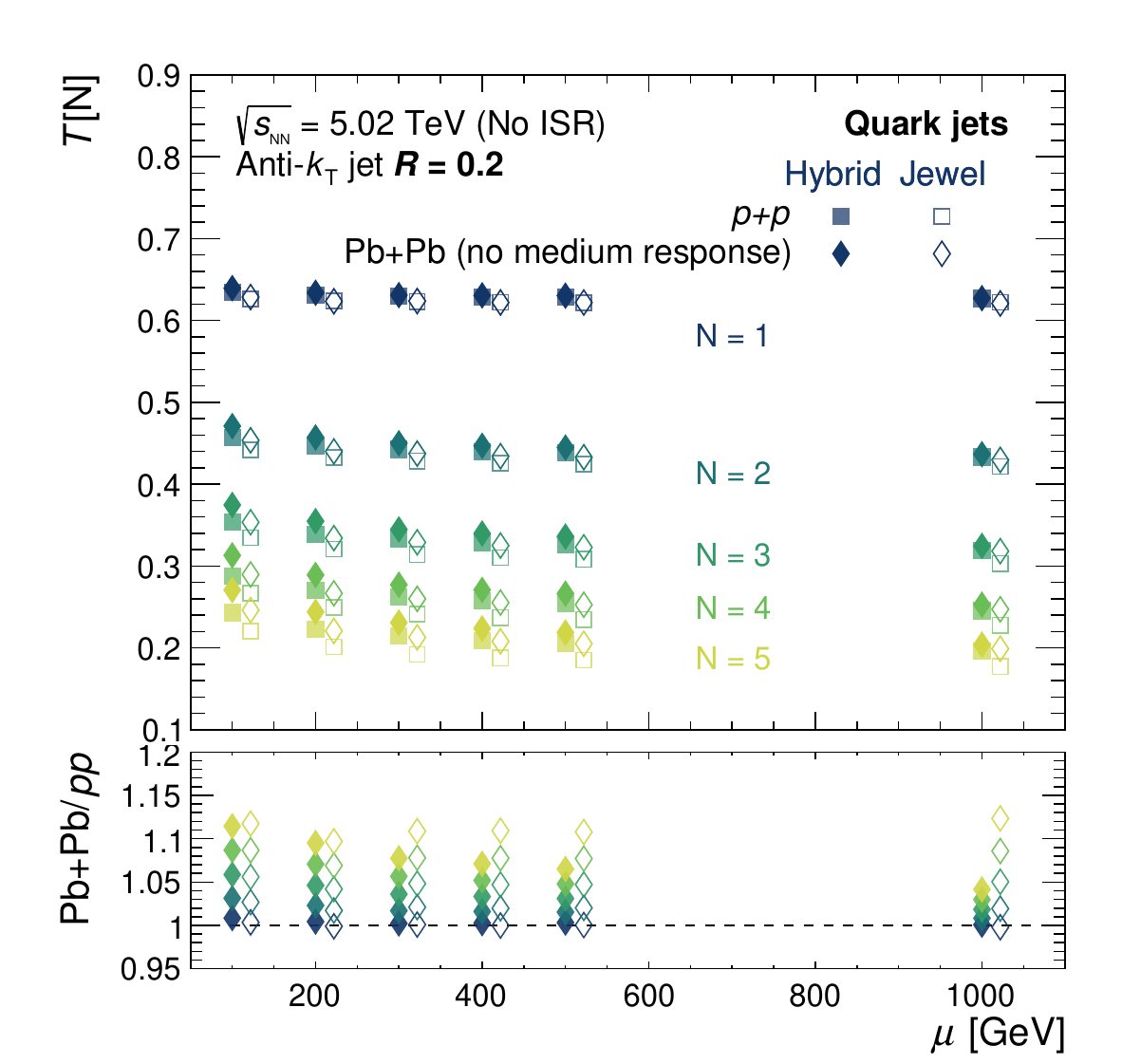}
    \includegraphics[width=0.48\textwidth]{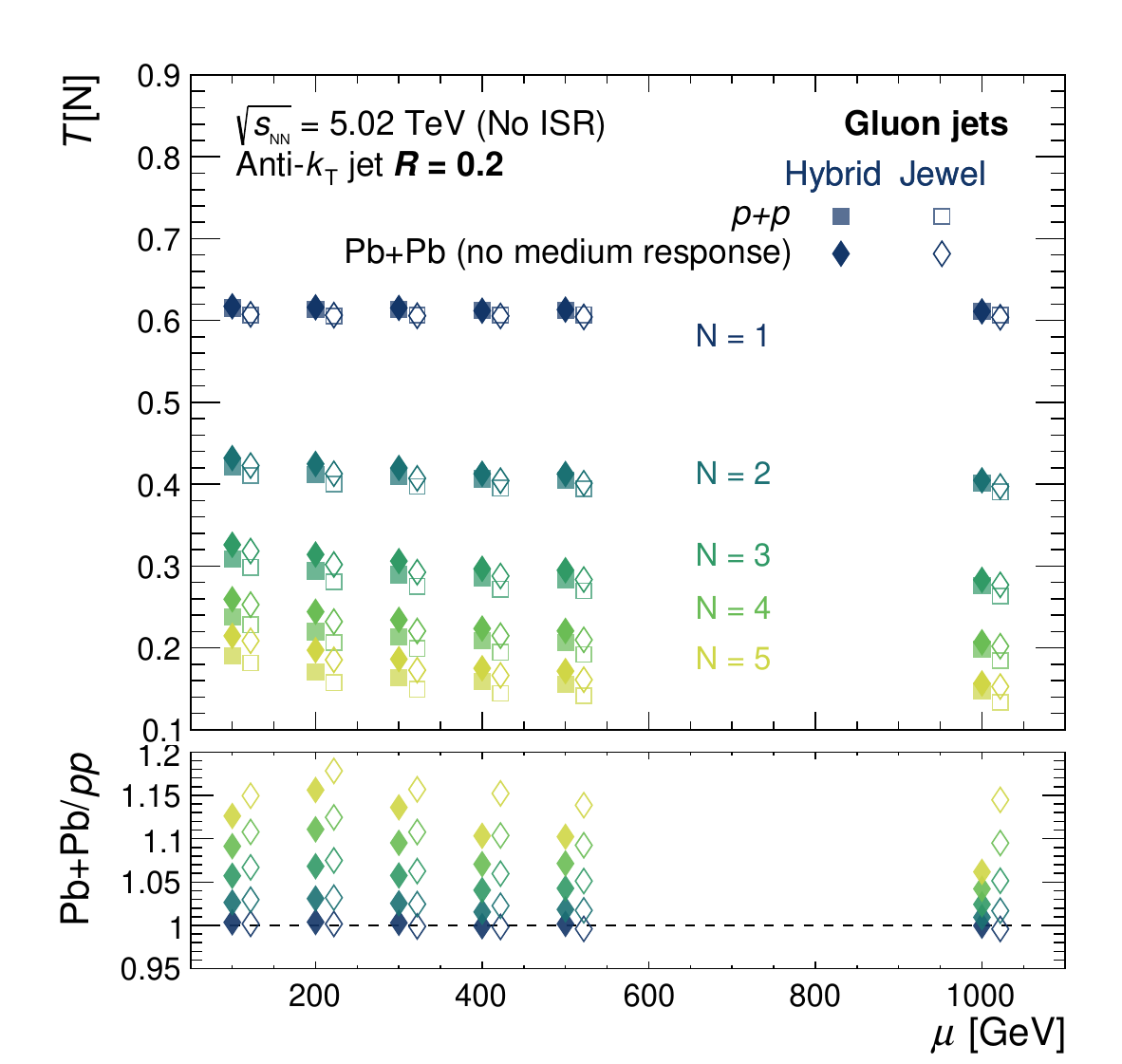}
    \caption{The N-th order moments of the track function as a function of $\mu$ for \pp collisions (squares) and \pbpb collisions (diamonds) without medium response in \hybrid (closed markers) and \jewel (open markers). The left panel is for quark-initiated jets and the right panel for gluon-initiated jet with $R$=0.2.
    The moment order $N$ increases from blue to yellow.
    The \jewel points are shifted slightly toward larger $\mu$ for visibility.
    The bottom panels represent the ratio of \pbpb to \pp.}
    \label{fig:jetTF_moment_vs_pthat_hybrid_jewel_R2}
\end{figure}

Figure~\ref{fig:jetTF_moment_vs_N_mu1000_R2} displays the moments for a fixed scale $\mu=1000$ GeV. To understand the dependence on $\mu$ of the track functions, we show in Fig.~\ref{fig:jetTF_moment_vs_pthat_hybrid_jewel_R2} the first through fifth moments of the track function as a function of the hard scale $\mu$ for \pp and \pbpb collisions, for quark-initiated and gluon-initiated jets with $R=0.2$. For a given $\mu$, the moments in \jewel are systematically slightly lower than those in \hybrid for both collision systems, indicating small but visible generator dependence already at the level of the average charged fraction and its low-order fluctuations. 

The bottom panels show the \pbpb to \pp ratios. Consistently with Fig.~\ref{fig:jetTF_moment_vs_N_mu1000_R2}, the modification is larger for higher moments at all $\mu$, reflecting that higher-order moments emphasize the tails of $T(x)$ where the \pbpb distribution is relatively enhanced. In \hybrid, the \pbpb to \pp ratios for the higher moments decrease with increasing $\mu$ and move closer to unity at $\mu=1000$~GeV. This behavior is consistent with a reduced effective interaction of harder jets with the medium; jets produced at larger $\mu$ are less susceptible to 
energy degradation, so the charged fraction distribution becomes more similar to its vacuum baseline. In contrast, \jewel exhibits a weaker $\mu$ dependence, and the higher-moment ratios remain significantly above unity across the full $\mu$ range, with a more pronounced enhancement for gluon-initiated jets. These trends highlight an interesting difference between both models in how the medium-induced modification of the charged fraction evolves with the hard scale, even when medium response is disabled. 
Eventually, as the medium scales become increasingly irrelevant at even higher $\mu$, one expects that even the highest moments collapse to the vacuum values in both jet quenching models. The fact that this asymptotic behaviour seems to be reached at different speeds for \jewel and \hybrid points to the different ways in which the same medium scales are imprinted in the evolution of the parton shower. The sensitivity of track functions to these modeling differences provides further motivation to study the scale dependence of the cumulants of track functions distributions, as we set out to do in the next subsection.

\begin{figure}[t!]
    \centering
    \includegraphics[width=0.48\linewidth]{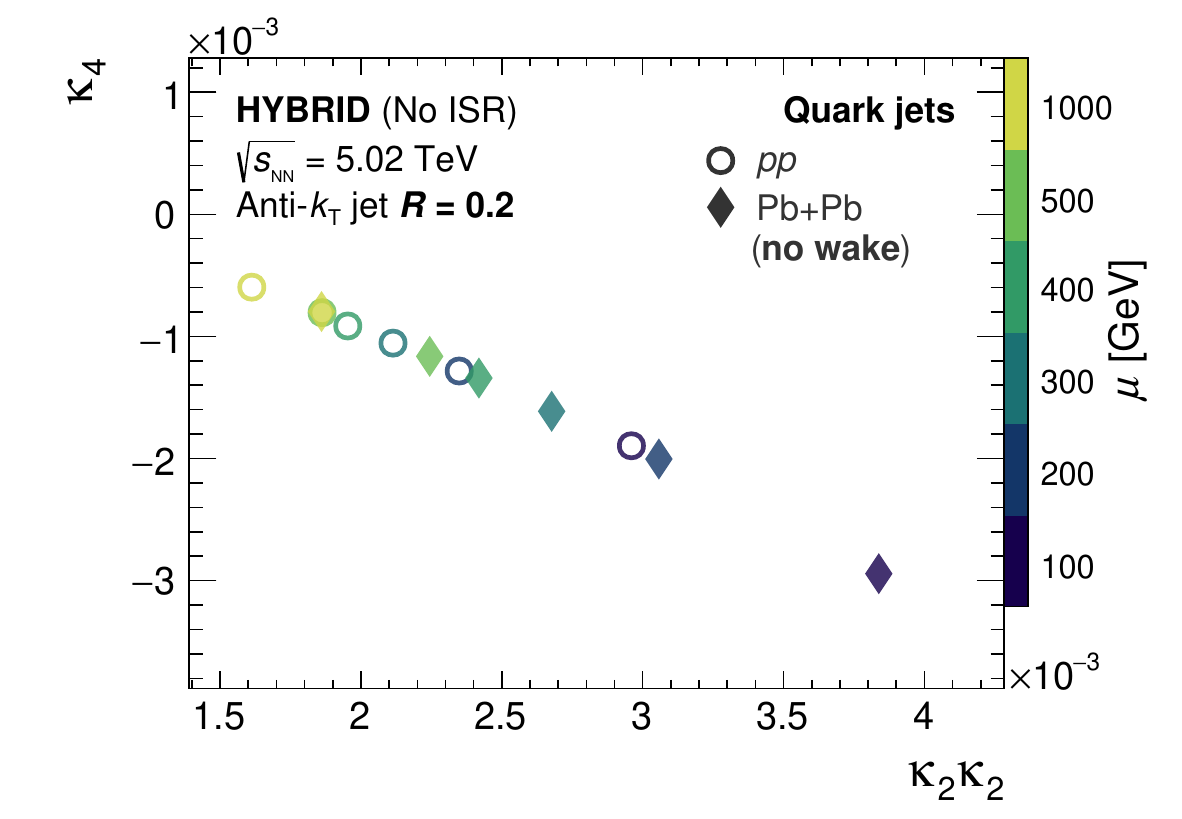}
    \includegraphics[width=0.48\linewidth]{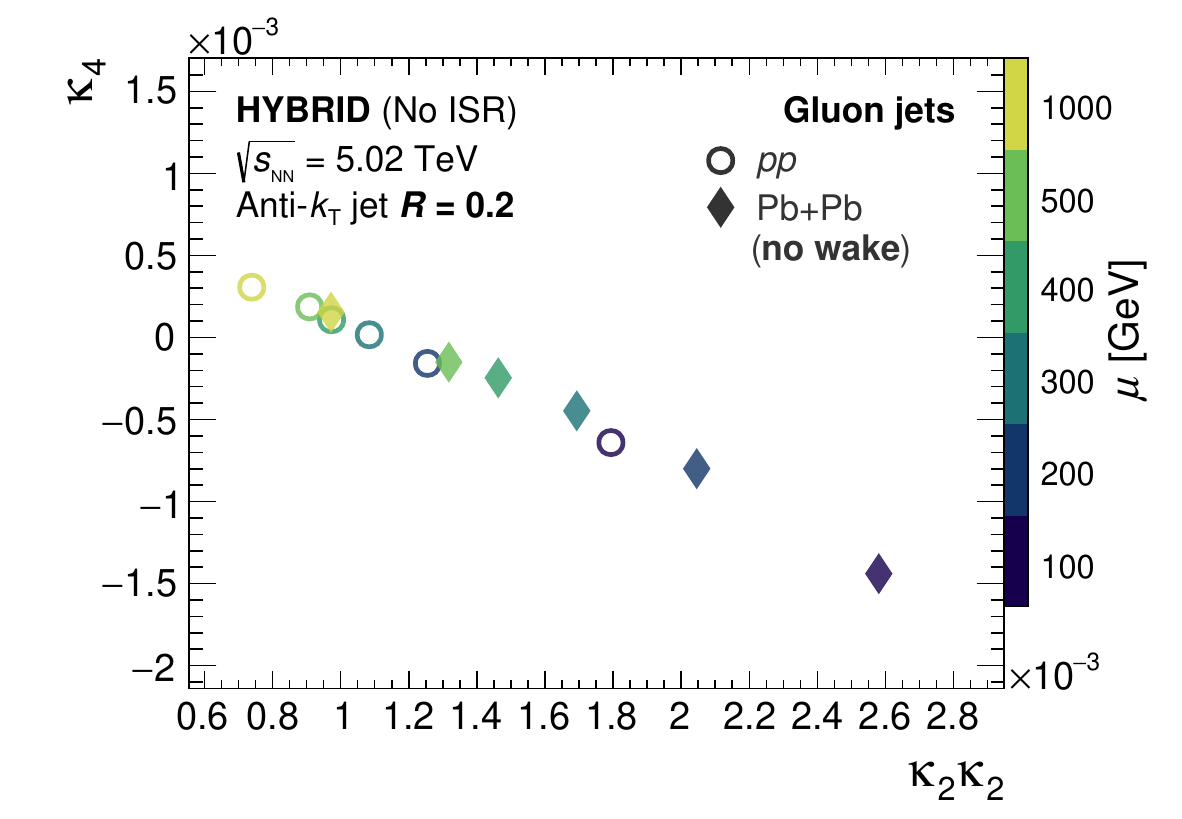}
    \includegraphics[width=0.48\linewidth]{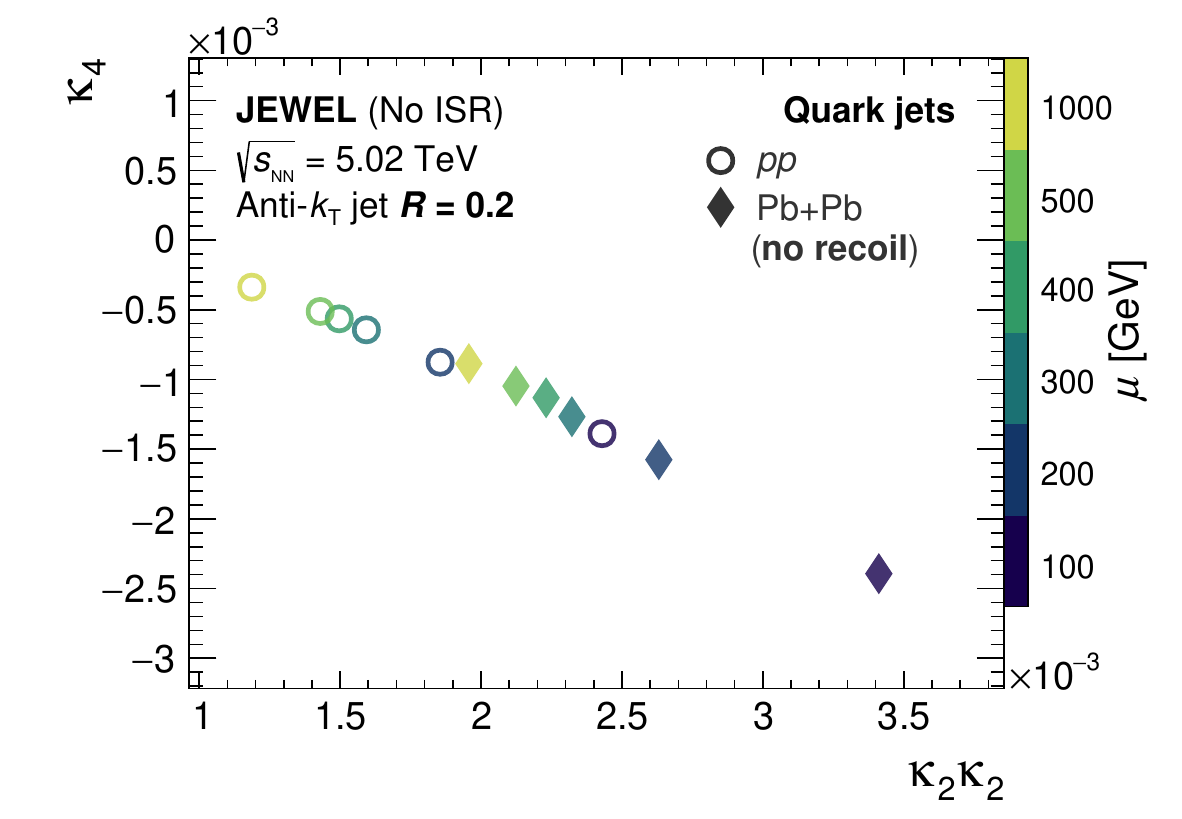}
    \includegraphics[width=0.48\linewidth]{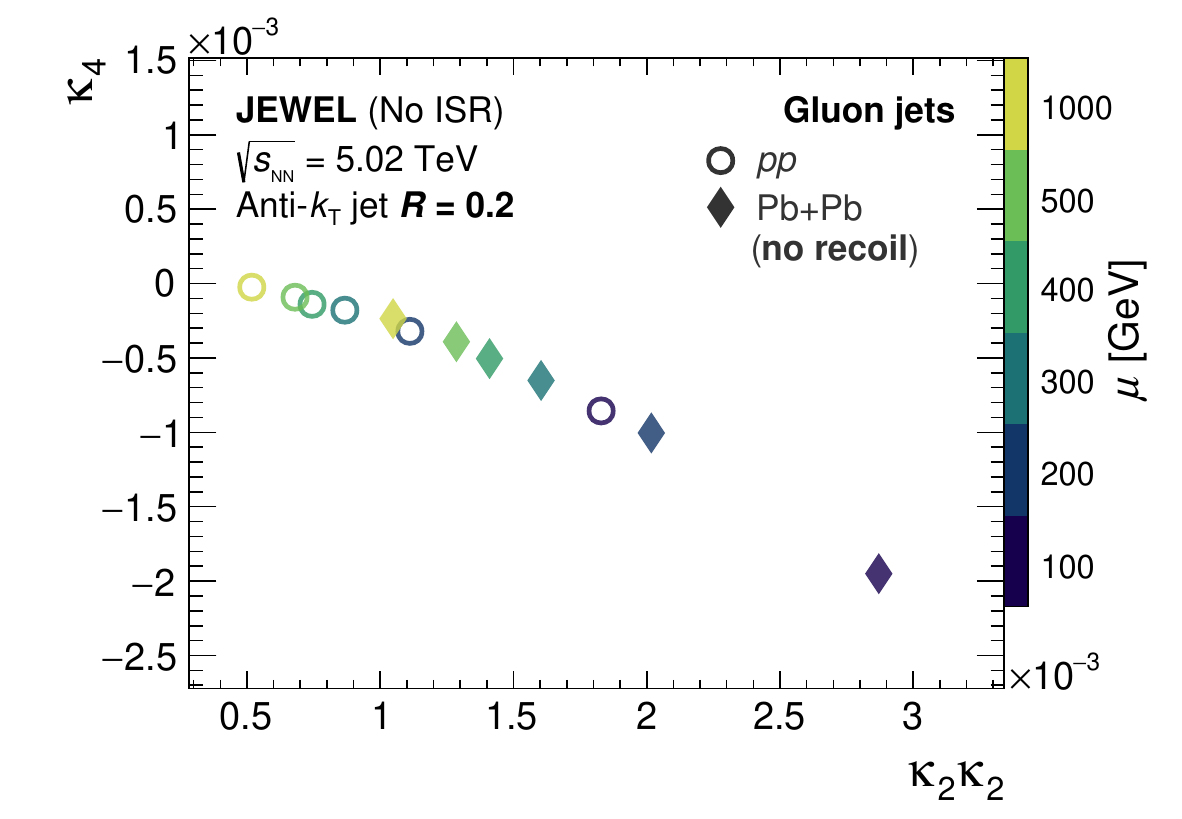}
    \caption{ $\kappa_4$ vs $\kappa_2^2$ cumulants of track functions for \pp (open circle) and \pbpb (filled diamond) without medium response. Results from the \hybrid (top) and \jewel (bottom) are shown for quark-initiated jets (left) and gluon-initiated jets (right) for $R$=0.2 jets.}
    \label{fig:jetTF_k4_vs_k22_R2}
\end{figure}

\subsection{Structure of the cumulants flow}

To characterize the shape of the track function distribution beyond its raw moments, we also consider the corresponding cumulants, which are computed from the central moments according to Equation~\ref{eq:cumdef}. 
Figure~\ref{fig:jetTF_k4_vs_k22_R2} 
shows the scale evolution in the ($\kappa_4,\kappa_2^2$) plane, 
with the color indicating the hard scale $\mu$, for quark- and gluon-initiated jets in \hybrid and \jewel. 
For each generator and jet flavor, the points trace a smooth, monotonic trajectory as $\mu$ varies. Increasing $\mu$ moves the points toward smaller $\kappa_2^2$ and toward $\kappa_4 \simeq 0$, indicating a progressive narrowing of the charged-fraction distribution and a reduction of its non-Gaussian cumulant component along the perturbative evolution. 
We note that in \jewel, at the highest value of $\mu$ explored, the PbPb point lies notably farther from its vacuum counterpart than it is the case for \hybrid. This is consistent with the observation made from Fig.~\ref{fig:jetTF_moment_vs_pthat_hybrid_jewel_R2} where we observed that the speed at which \jewel convergences to the vacuum value is slower than \hybrid's.

The evolution in \pp collisions is consistent with the structure of Eqs.~\eqref{eq:kappa2RG}, see e.g.~\cite{Lee:2023xzv}. Remarkably, the \pbpb points lie along the same trajectory as the \pp ones, albeit with a different initial condition for both sets. From Eqs.~\eqref{eq:kappa2RG}, the fact that the \pbpb points lie along the same trajectory as
the \pp ones suggests that the relative structure of the cumulant flow is largely unmodified, as any (inhomogenous) rescaling would lead to a different trajectory in the ($\kappa_4$,$\kappa_2^2$) plane. This observation is noteworthy because it indicates that, within these MC models, the scale evolution of in-medium jets is largely governed by the same perturbative dynamics as in vacuum, with the medium primarily modifying the effective boundary conditions of the shower.
This is consistent with the theoretical expectation that the interactions with the background medium do not modify the scale evolution of the track functions. If observed experimentally, such a pattern would provide strong evidence that the virtuality cascade of in-medium jets develops essentially as in vacuum. This assumption lies at the heart of several jet quenching models~\cite{Schenke:2009gb,Casalderrey-Solana:2014bpa,He:2015pra,Caucal:2019uvr,Mehtar-Tani:2021fud,Takacs:2021bpv,Apolinario:2026hff}, but has never been directly quantified; a measurement of the type shown in Fig.~\ref{fig:jetTF_k4_vs_k22_R2} offers a promising path towards addressing this question.

\begin{figure}[t!]
    \centering
       \includegraphics[width=0.68\linewidth]{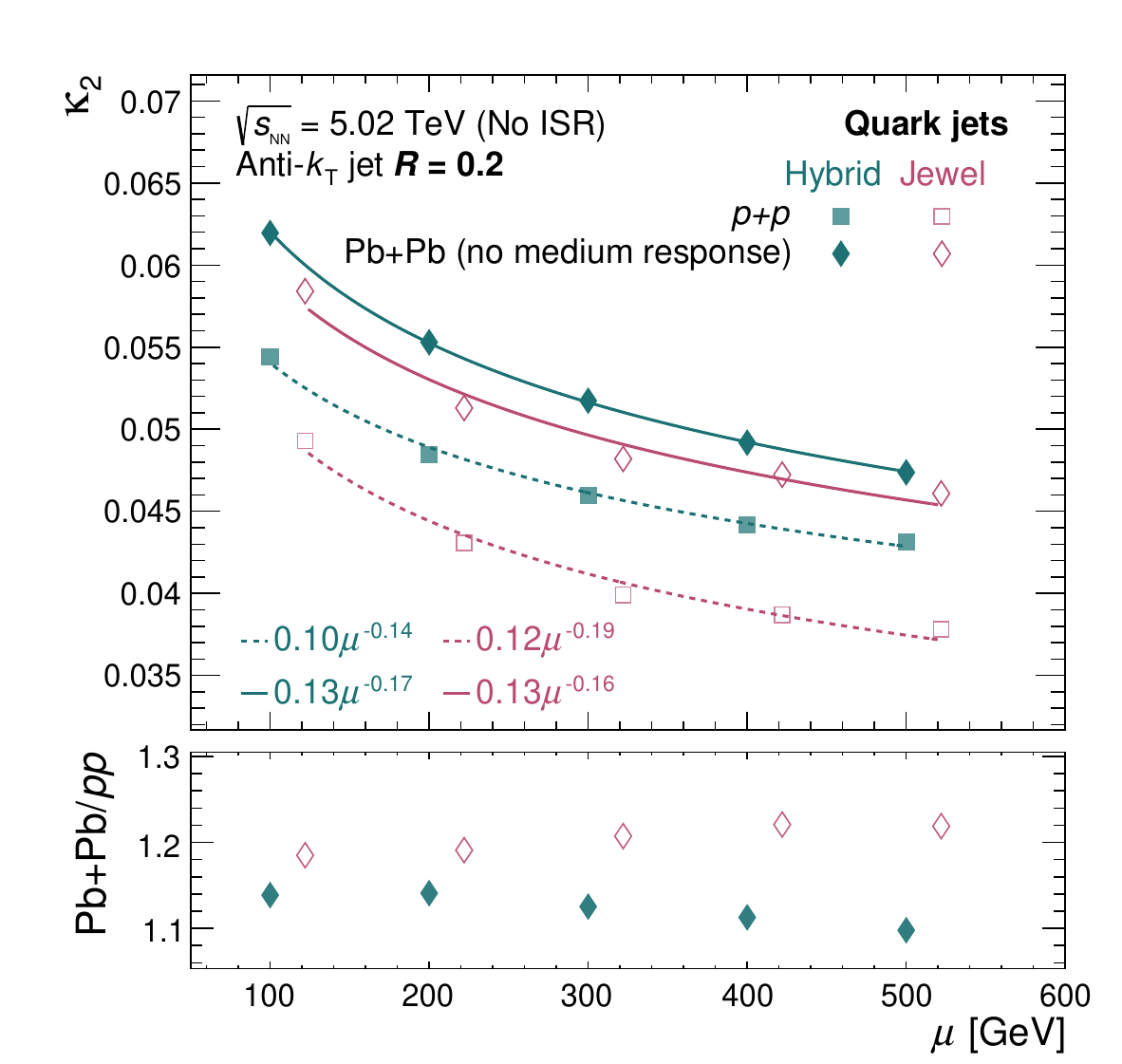}
    \caption{ $\kappa_2$ cumulant as a function of $\mu$ for quark-initiated jets with $R$=0.2 in \pp (squares) and \pbpb (diamonds) collisions without medium response for \hybrid (green closed markers) and \jewel (pink open markers). The lines represent power-law fits. The \jewel points are shifted slightly toward larger $\mu$ for visibility. The bottom panel shows the corresponding \pbpb to \pp ratio.}
    \label{fig:kappa2_vs_mu}
\end{figure}

Although Fig.~\ref{fig:jetTF_k4_vs_k22_R2} strongly indicates that the RG flow is not deeply modified, we can further explore this point in Fig.~\ref{fig:kappa2_vs_mu}, where we provide the evolution of $\kappa_2$ with $\mu$ for quark-initiated jets. The data points are fitted to a power law decay, motivated by the analytic solution of Eqs.~\eqref{eq:kappa2RG} in the case of a purely gluonic theory; the full QCD solution is more involved but the qualitative behavior should persist. As shown in the ratio, the dependence in $\mu$ is very weak. In the case of the gluon-only evolution, this would indicate that the anomalous scaling in \pbpb collisions is consistent with that in \pp collisions. In other words, this would imply that the virtuality cascade in both systems has the exact same evolution and only differs (minimally) by boundary effects or due to different initial conditions. 
Although the extension of this interpretation to full QCD is not immediate, the qualitative conclusion remains that the medium does not significantly modify the scale dependence of the cumulant evolution.

\begin{figure}[t!]
    \centering
       \includegraphics[width=0.48\linewidth]{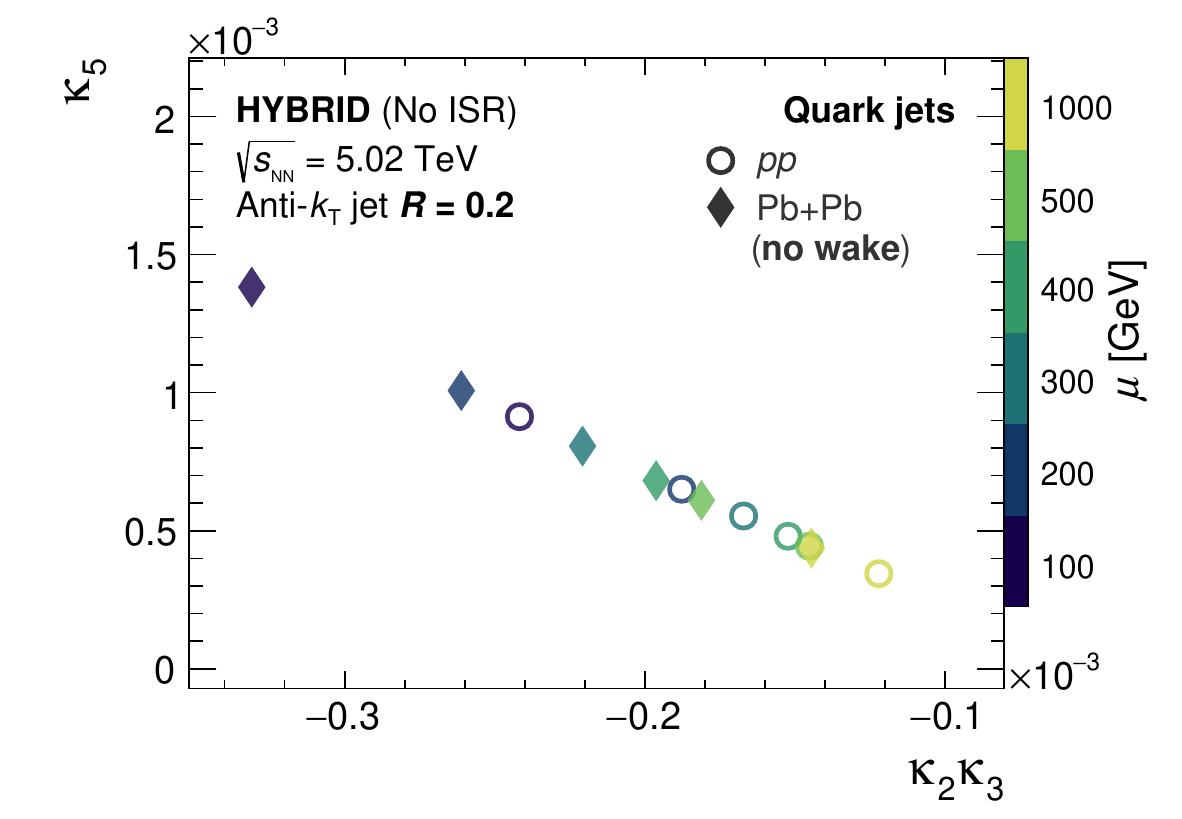}
    \includegraphics[width=0.48\linewidth]{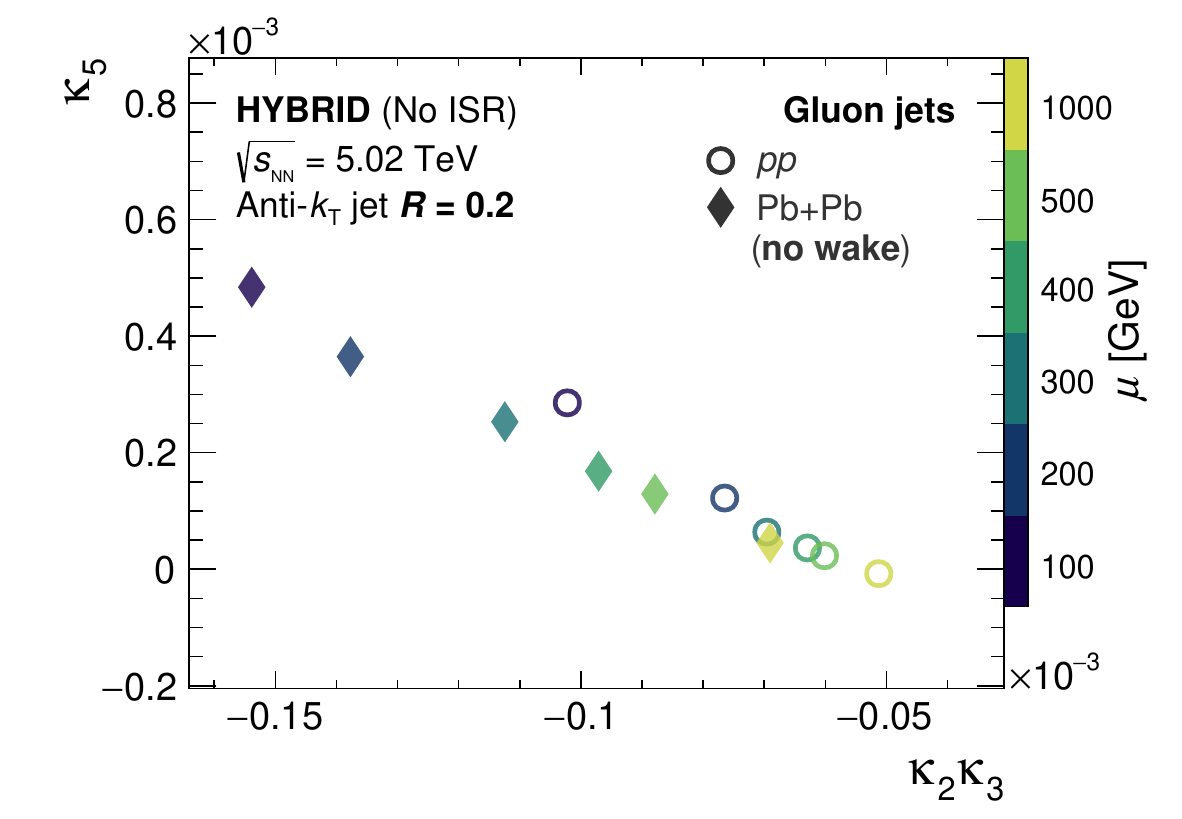}
    \includegraphics[width=0.48\linewidth]{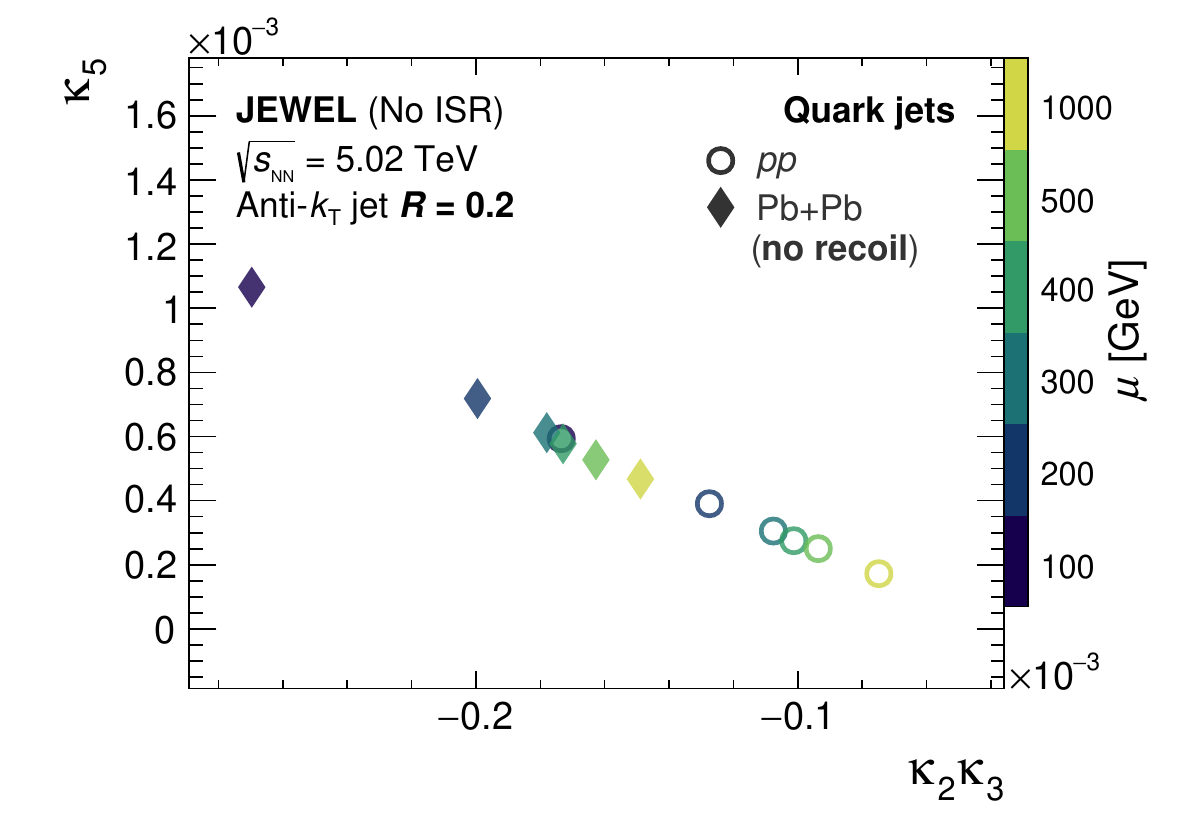}
    \includegraphics[width=0.48\linewidth]{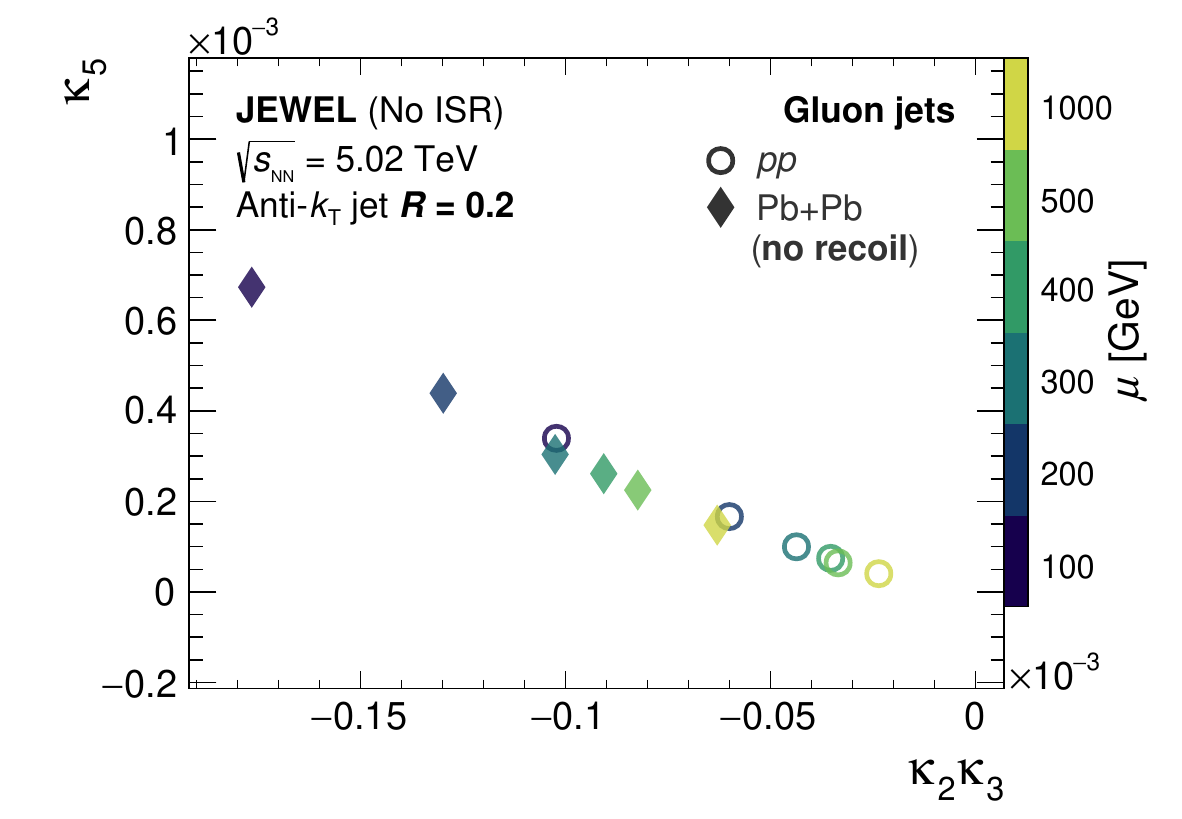}
    \caption{ $\kappa_5$ vs $\kappa_2\kappa_3$ cumulants of track functions for \pp (open circle) and \pbpb (filled diamond). Results from the \hybrid (top) and \jewel (bottom) are shown for quark-initiated jets (left) and gluon-initiated jets (right) for $R$=0.2 jets. 
    }
    \label{fig:jet_tf_k5_vs_k2k3_R2}
\end{figure}

Finally, Figure~\ref{fig:jet_tf_k5_vs_k2k3_R2} extends this analysis to higher cumulants. The \pp and \pbpb trajectories remain nearly aligned, indicating that the qualitative behavior discussed above persists beyond the lowest cumulants. We note, however, that the evolution involving $\kappa_6$, shown in the Appendix, exhibits larger deviations. Since higher cumulants are more sensitive to the tails of the track function and to statistical fluctuations, the present analysis does not allow us to determine whether these deviations reflect genuine physical effects or are driven by limited statistics.

\subsection{Experimental considerations}

The results presented here provide a useful basis for assessing the experimental feasibility of track function measurements in heavy-ion collisions. A central experimental challenge is the separation of genuine modifications of the perturbative jet cascade from contributions associated with medium response and the underlying event. As shown in the Appendix, medium response in both \hybrid and \jewel can substantially affect the cumulant evolution, particularly for larger jet radii. This effect is reduced for narrow jets, such as $R=0.2$, for which the medium response contribution has limited overlap with the reconstructed jet cone. This suggests that measurements with small-radius jets provide a cleaner experimental configuration for isolating modifications of the hard shower.

Another relevant consideration is the experimental reconstruction of low-\pT charged particles. To assess the robustness of the proposed observables against realistic tracking thresholds, we repeated the analysis by requiring charged particles to satisfy $\pT>0.5~\mathrm{GeV}$. The resulting track functions, moments, and cumulants are found to be nearly unchanged with respect to the nominal results. This indicates that the observables considered here are not dominated by extremely soft particles and should remain experimentally accessible with standard charged-particle reconstruction capabilities.

Measurements in smaller collision systems~\cite{Grosse-Oetringhaus:2024bwr}, such as oxygen-oxygen collisions, could provide a particularly useful intermediate step between \pp and \pbpb collisions. Such systems would allow one to study the onset of medium-induced modifications in an environment with reduced underlying-event activity compared to \pbpb collisions. Measurements at different collision energies would also be valuable. In particular, lower-energy heavy-ion data, such as those accessible at RHIC, would extend the study to lower jet transverse momenta and hence lower effective evolution scales, where the present results indicate that medium-induced deviations from the vacuum baseline are more pronounced. Taken together, measurements across system size, collision energy, and jet radius would provide a direct experimental test of whether the scale evolution of track functions remains vacuum-like in the presence of a QGP.

\section{Conclusions}\label{sec:conclusion}
In this work we have provided the first phenomenological study of track functions in heavy-ion collisions. Our study motivates the application of these objects as optimal probes of the in-medium jet cascade. Since the track functions are sensitive to the entire fragmentation process, unlike the more commonly studied fragmentation functions, they allow to probe the entire correlations built during the branching process in matter. As their RG flow is well understood in vacuum, this allows to perform direct tests of the jet fragmentation which can be compared to perturbative QCD expectations. This feature is seldomly found in the jet quenching literature, and offers new opportunities for more precise understanding of jet fragmentation in the QGP. 

The numerical results presented above substantiate this picture in a concrete way. Already with medium response switched off, the distributions in heavy-ion collisions differ from their proton-proton collision baselines most clearly in higher moments and cumulants, showing that track functions are sensitive to changes in the internal structure of the cascade rather than only to the overall jet energy loss. 
This effect is observed in both \hybrid and \jewel, although with different magnitudes and scale dependence, indicating that track functions can help discriminate between microscopic descriptions of jet--medium interactions. At the same time, the cumulant RG flows in heavy-ion collisions remain closely aligned with the corresponding \pp collision trajectories, suggesting that the perturbative part of the virtuality cascade is largely vacuum-like, while the medium mainly modifies the effective boundary conditions of the shower. 
Together, these features make track functions a promising target for future experimental studies.

\acknowledgments
DP is supported by the Spanish Ram\'on y Cajal fellowship RYC2023-044989-I.

\bibliographystyle{jhep}
\bibliography{cite.bib}

\newpage
\appendix
\section*{Appendix}
\input{appendix}

\end{document}

%% file: appendix.tex
\label{app:temp}


The purpose of this Appendix is to complement the material provided in the main text by varying the scale $\mu$, the jet radius $R$, and turning on and off medium response. 

\section{Impact of Jet Radius on Track Function Distribution}
\begin{figure}[h!]
\centering    
    \includegraphics[width=0.48\textwidth]{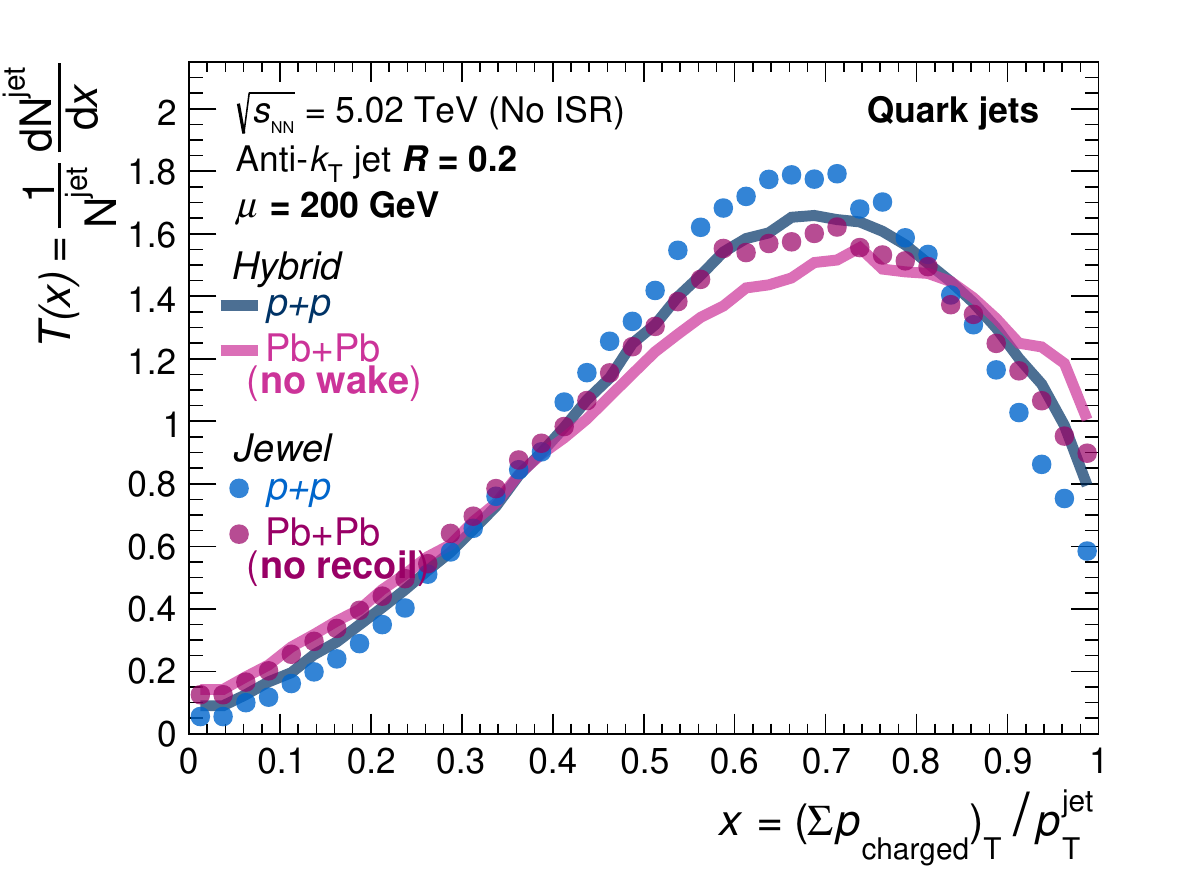}
    \includegraphics[width=0.48\textwidth]{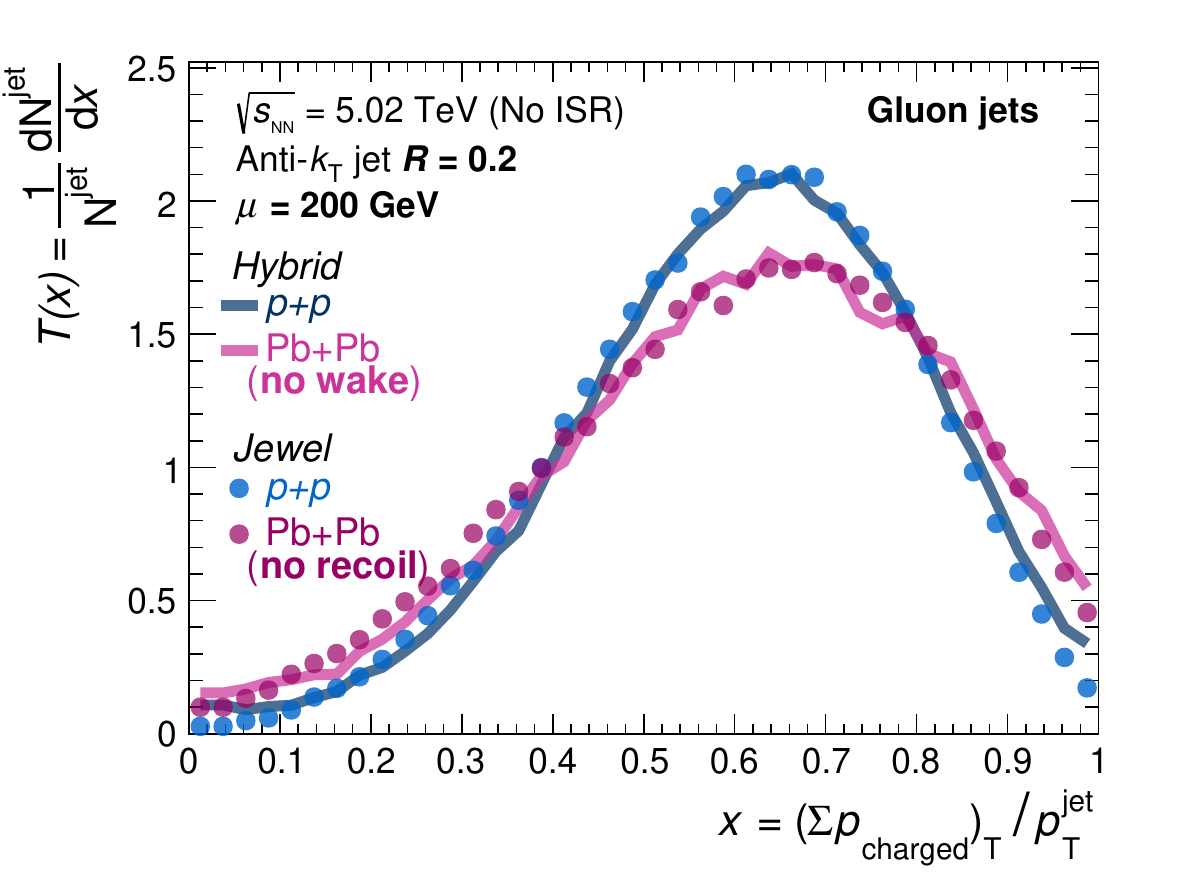}
    \caption{Track function for quark-initiated (left) and gluon-initiated jets (right) for \hybrid (lines) and \jewel (markers) in \pp (blue) and \pbpb (pink) collisions without medium response. Jets with $R$=0.2 and the $\mu$ scale is 200 GeV}.
    \label{fig:jetTF_hybrid_jewel_pthat200_R2}
\end{figure}

Figure~\ref{fig:jetTF_hybrid_jewel_pthat200_R2} shows the track functions at $\mu=200$ GeV for $R=0.2$ jets in \pp and \pbpb collisions without medium response. For both quark-initiated and gluon-initiated jets, the distributions are centered around the expected charged fraction, $x \sim 0.6-0.7$. Compared to \pp collisions, the \pbpb distributions are generally broader and show a reduced peak height, indicating an increased event-by-event fluctuation of the charged momentum fraction in the presence of the medium. This broadening is observed in both \hybrid and \jewel, although the quantitative size of the modification differs between the two generators. The effect is more pronounced for gluon-initiated jets, consistent with their larger color charge and stronger sensitivity to medium-induced modifications of the shower. Compared to the larger hard scale $\mu=1000$ GeV shown in Fig.~\ref{fig:jetTF_hybrid_jewel_pthat1000_R2}, the differences between \pp and \pbpb are larger, as expected, since larger $\mu$ means a larger separation between the vacuum and medium scales. Note that while the size of medium effects is comparable between \jewel and \hybrid for $\mu=200$ GeV, that is not the case for the scale $\mu=1000$ GeV shown in the main text since \jewel's asymptotic approach to the vacuum result is notably slower than \hybrid's. 

\begin{figure}[ht!]
\centering    
    \includegraphics[width=0.48\textwidth]{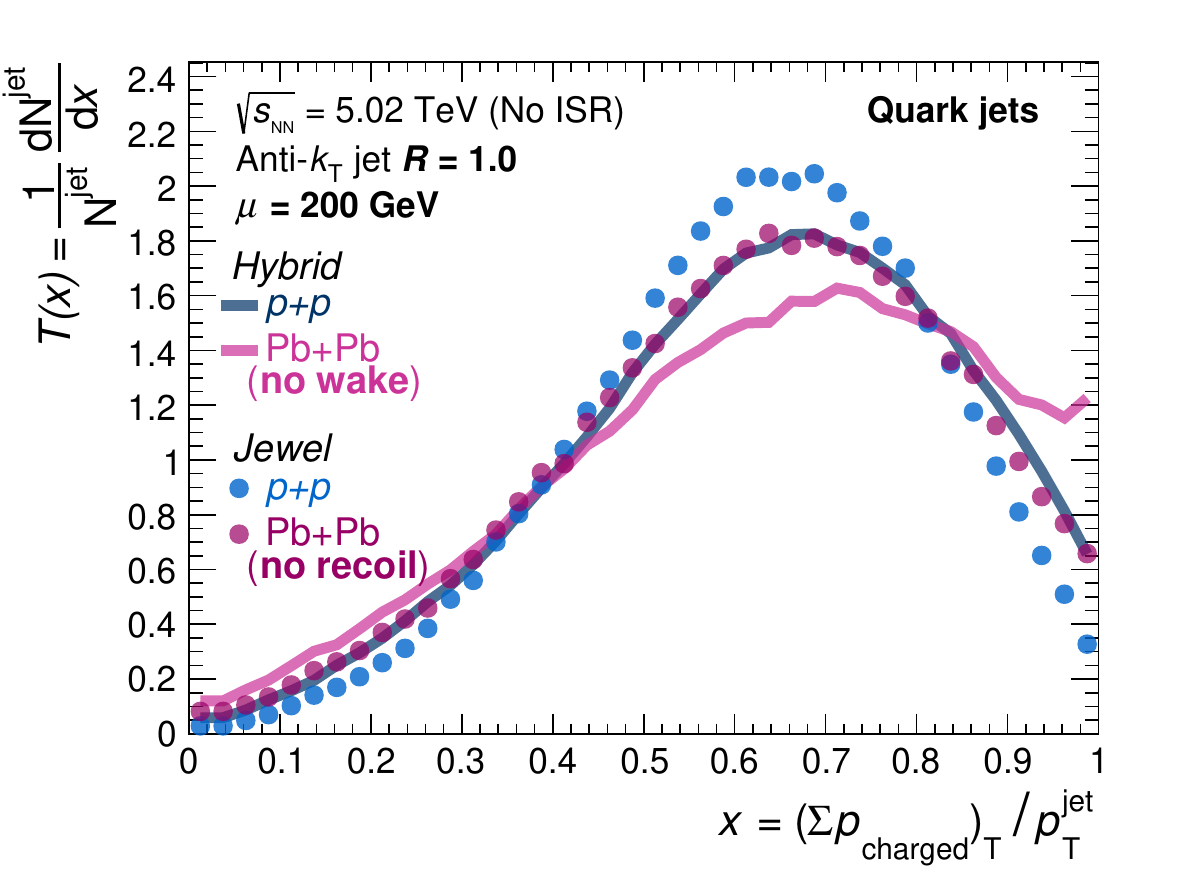}
    \includegraphics[width=0.48\textwidth]{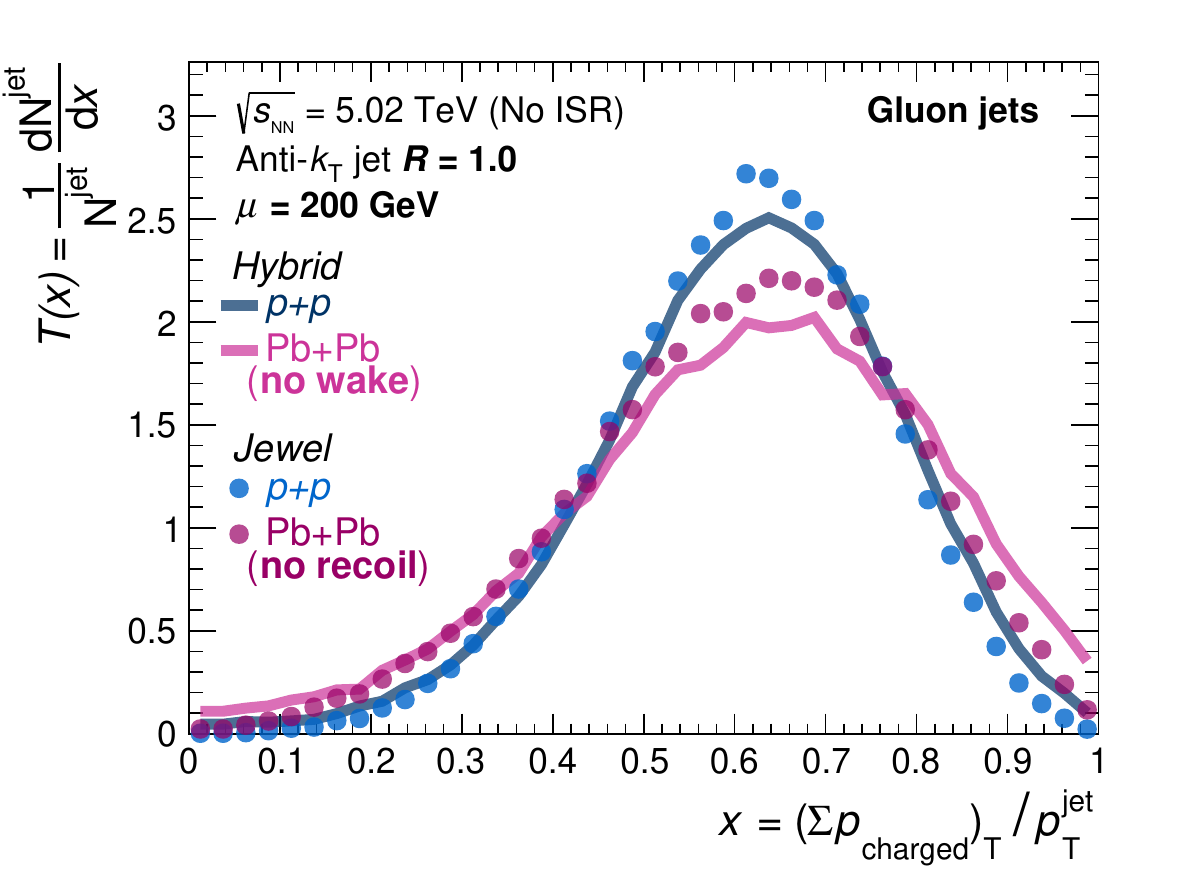}
    \caption{Track function for quark-initiated (left) and gluon-initiated jets (right) for \hybrid (lines) and \jewel (markers) in \pp and \pbpb collisions without medium response. Jets with $R$=1.0 and the $\mu$ scale is 200 GeV.}
    \label{fig:jetTF_hybrid_jewel_pthat200_R10}
\end{figure}

\begin{figure}[h!]
\centering    
    \includegraphics[width=0.48\textwidth]{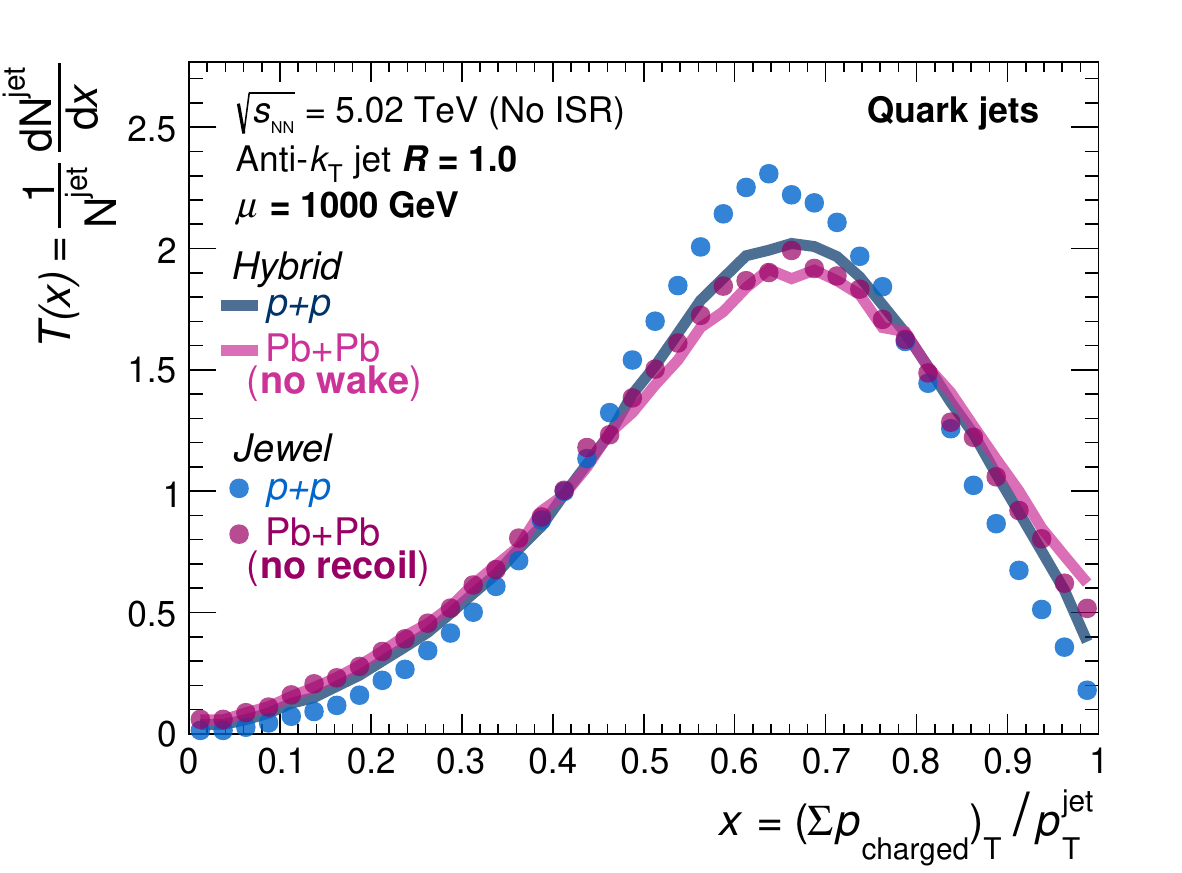}
    \includegraphics[width=0.48\textwidth]{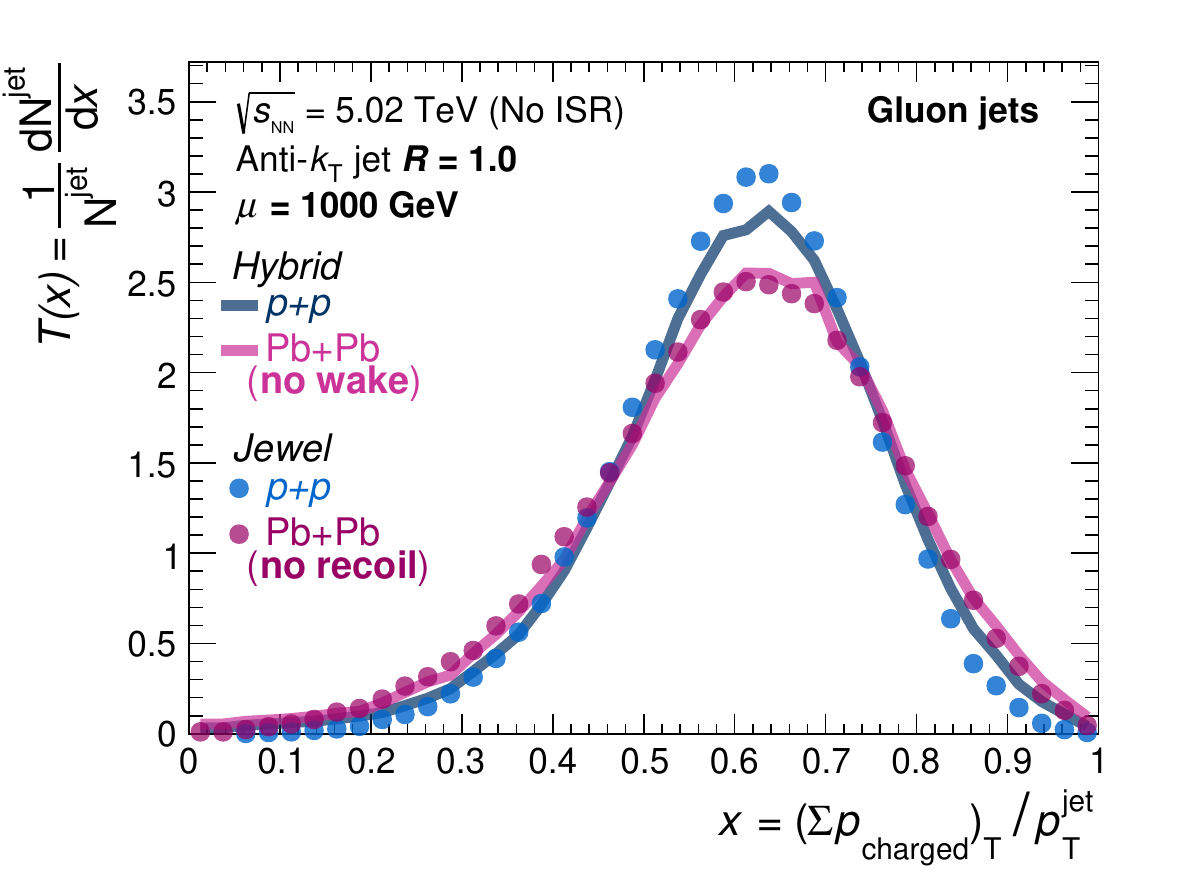}
    \caption{Track function for quark-initiated (left) and gluon-initiated jets (right) for \hybrid (lines) and \jewel (markers) in \pp and \pbpb collisions without medium response. Jets with $R$=1.0 and the $\mu$ scale is 1000 GeV}.
    \label{fig:jetTF_hybrid_jewel_pthat1000_R10}
\end{figure}

The track function distributions for $R=1.0$ jets are shown at $\mu=200$ GeV in Fig.~\ref{fig:jetTF_hybrid_jewel_pthat200_R10} and at $\mu=1000$ GeV in Fig.~\ref{fig:jetTF_hybrid_jewel_pthat1000_R10}.
The qualitative behavior remains the same as for $R=0.2$ jets. Given the challenge to deal with large radius jets due to the large background present in heavy-ion collisions, for the physics we pursue in this paper, the choice of $R=0.2$ is preferred over larger $R$ values. In any case, a noteworthy difference between small and large radius is that, as one would expect, the larger multiplicity contained in wider cones has the effect of narrowing the track function distribution.


\clearpage
\section{Medium Response Effect on Track function Distribution}

\begin{figure}[hbtp!]
\centering    
    \includegraphics[width=0.48\textwidth]{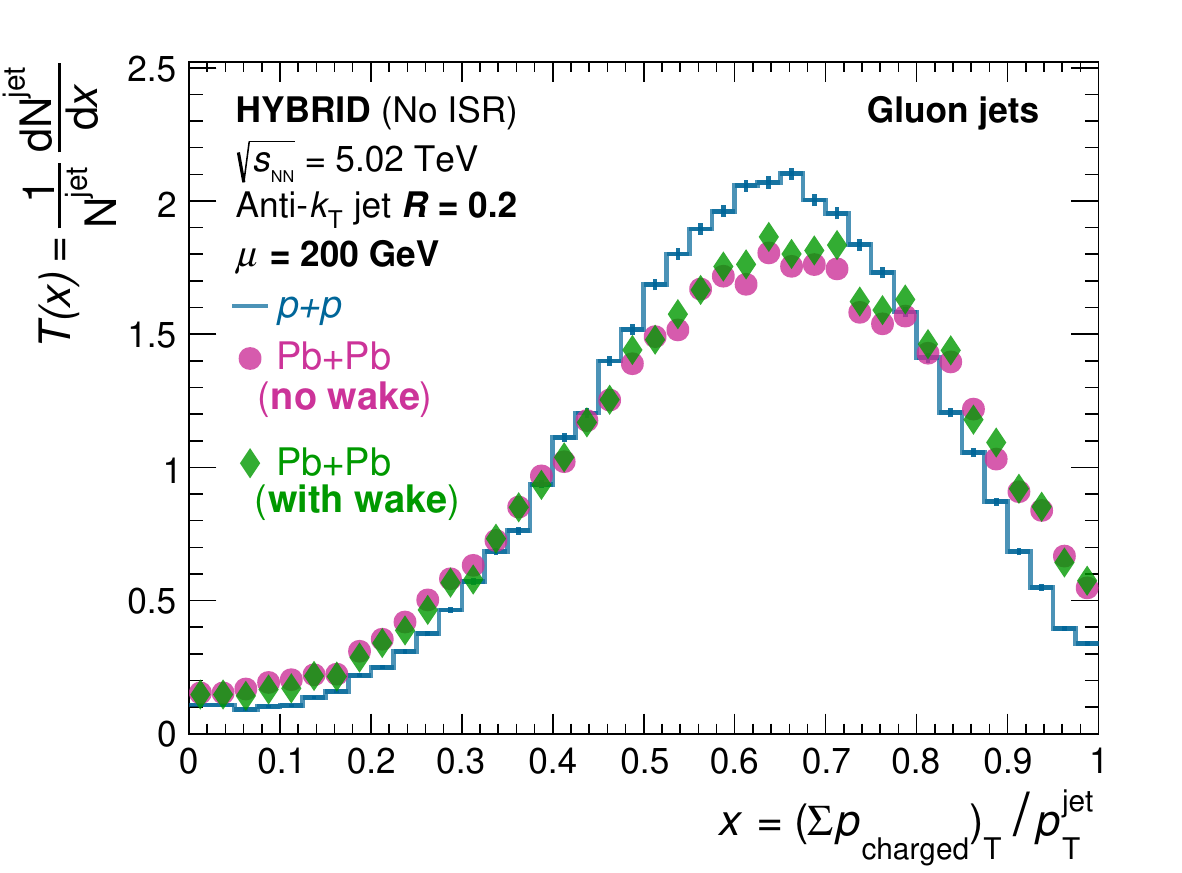}
    \includegraphics[width=0.48\textwidth]{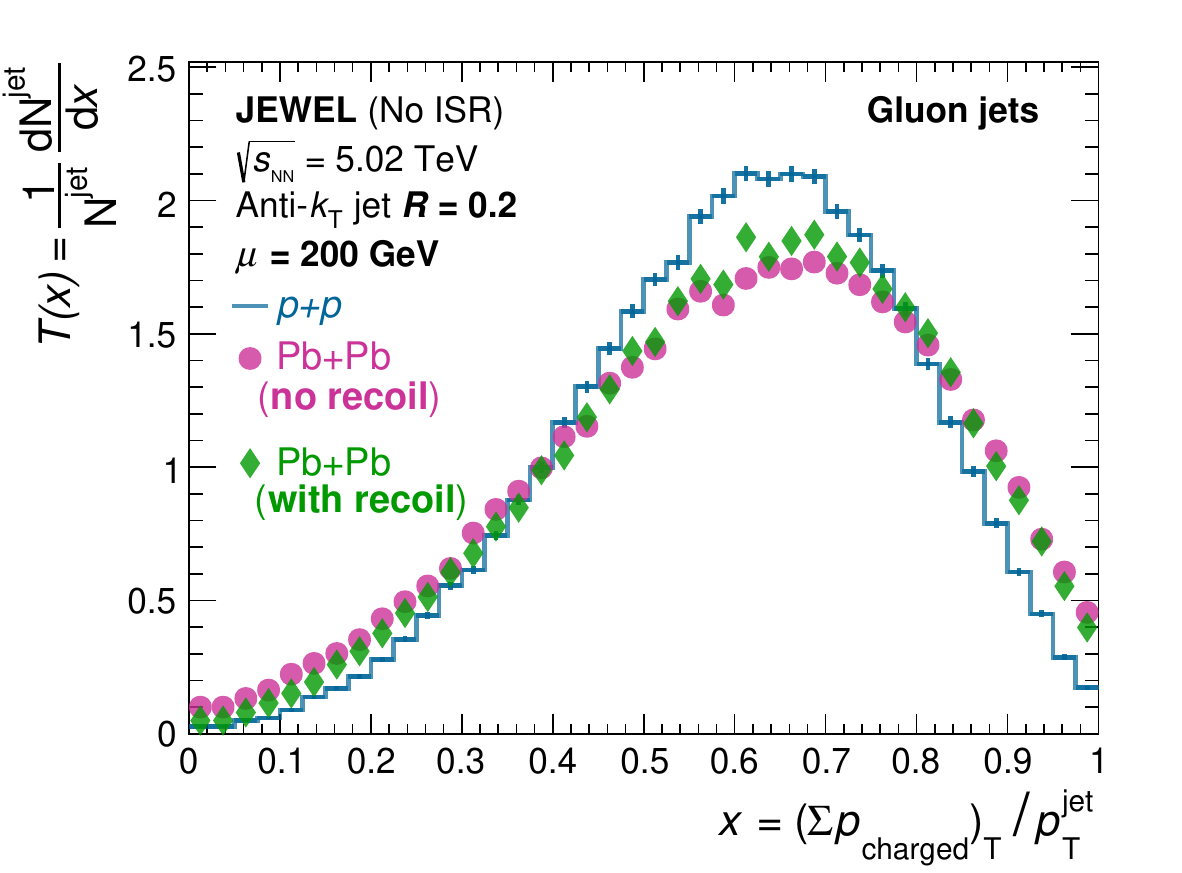}
    \includegraphics[width=0.48\textwidth]{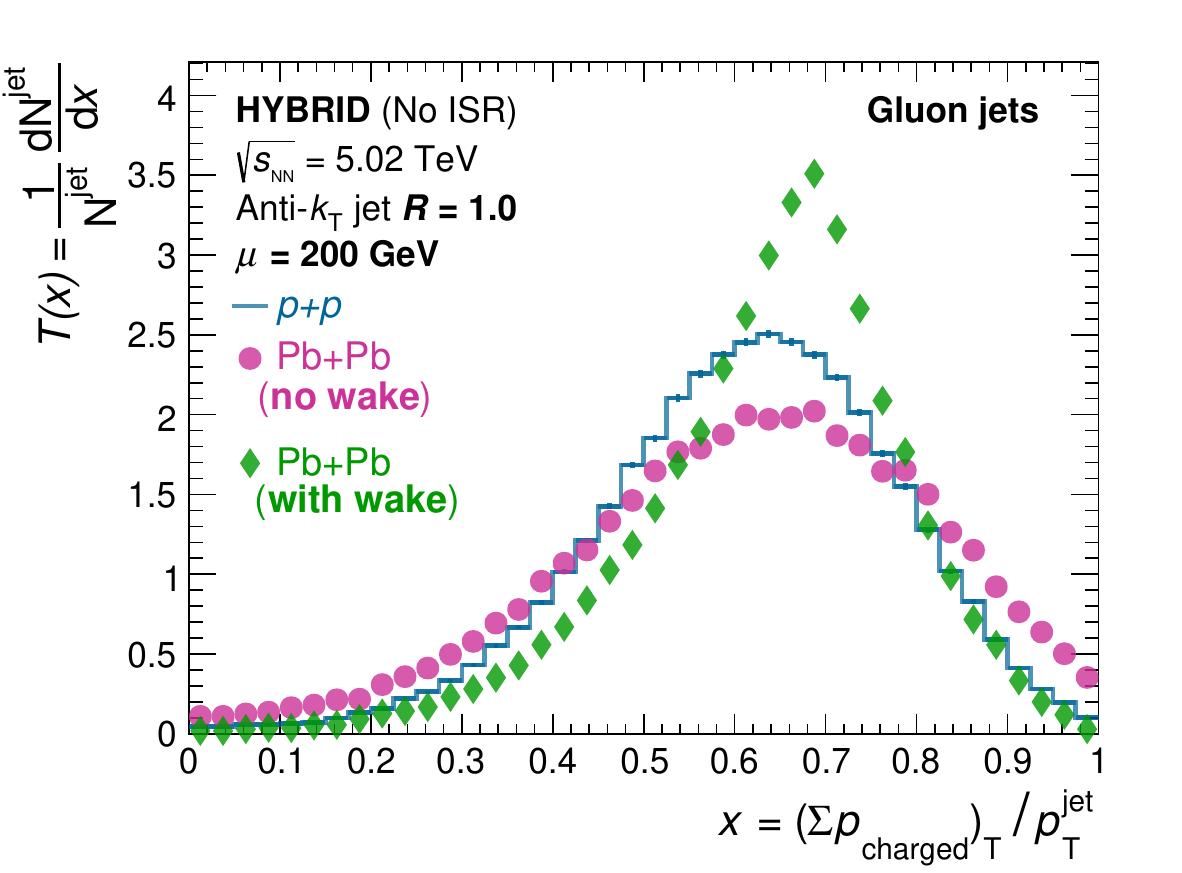}
    \includegraphics[width=0.48\textwidth]{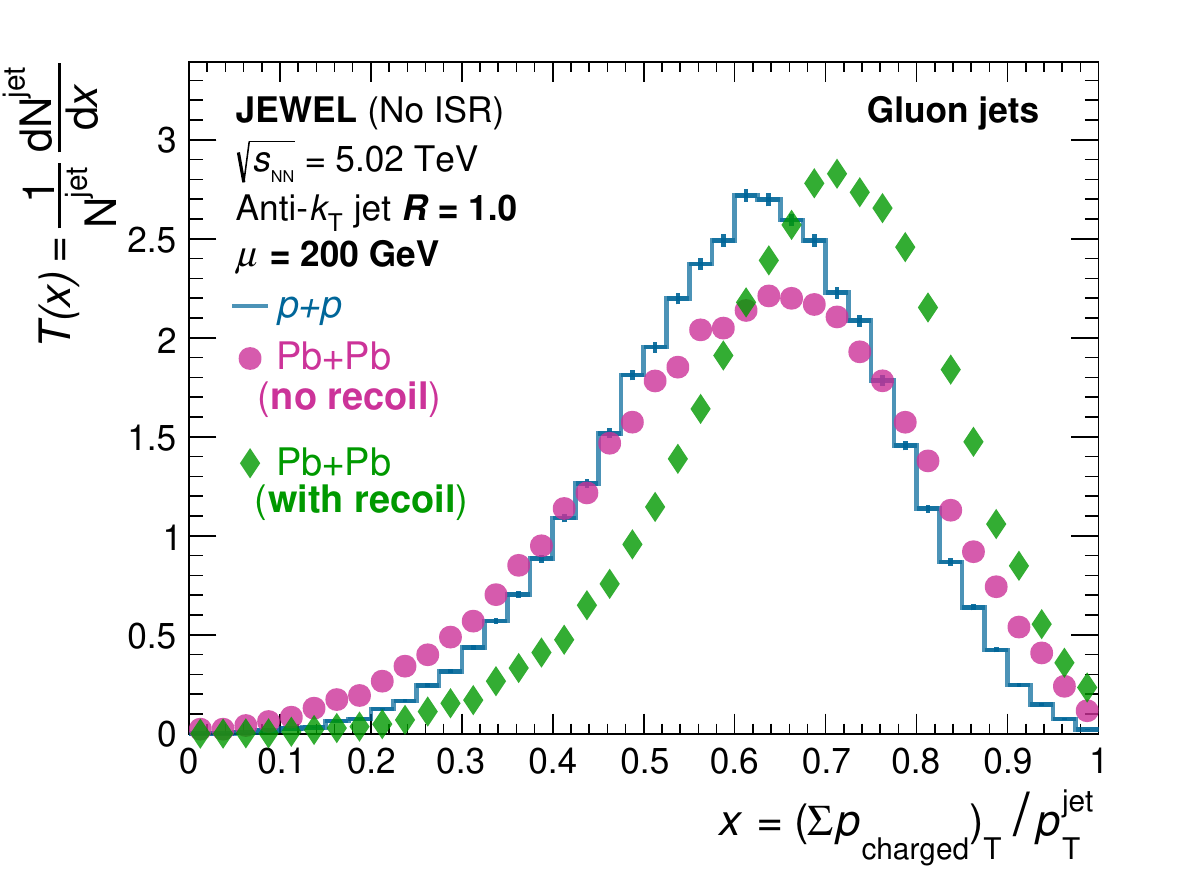}
    \caption{
    Track functions for gluon-initiated jets showing the impact of medium response at $\mu=200~\mathrm{GeV}$. Results are shown for \hybrid (left panels) and \jewel (right panels), comparing the \pp baseline (blue), \pbpb without medium response (pink circles), and \pbpb with medium response (green diamonds).
    Medium response corresponds to wake contributions in \hybrid and recoil contributions in \jewel. The top and bottom panels show $R=0.2$ and $R=1.0$ jets, respectively.}
    \label{fig:jetTF_hybrid_jewel_pthat200_wakeComp}
\end{figure}

\begin{figure}[hbtp!]
\centering    
    \includegraphics[width=0.48\textwidth]{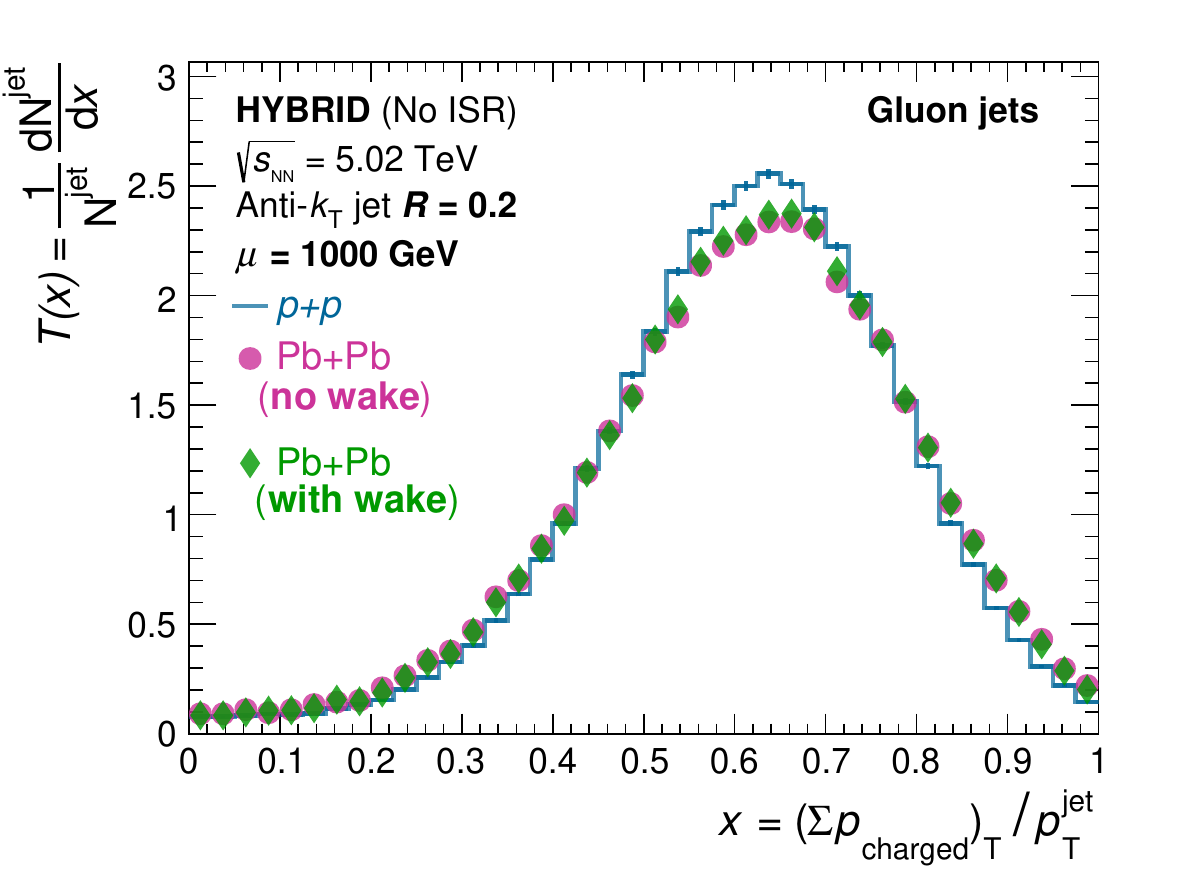}
    \includegraphics[width=0.48\textwidth]{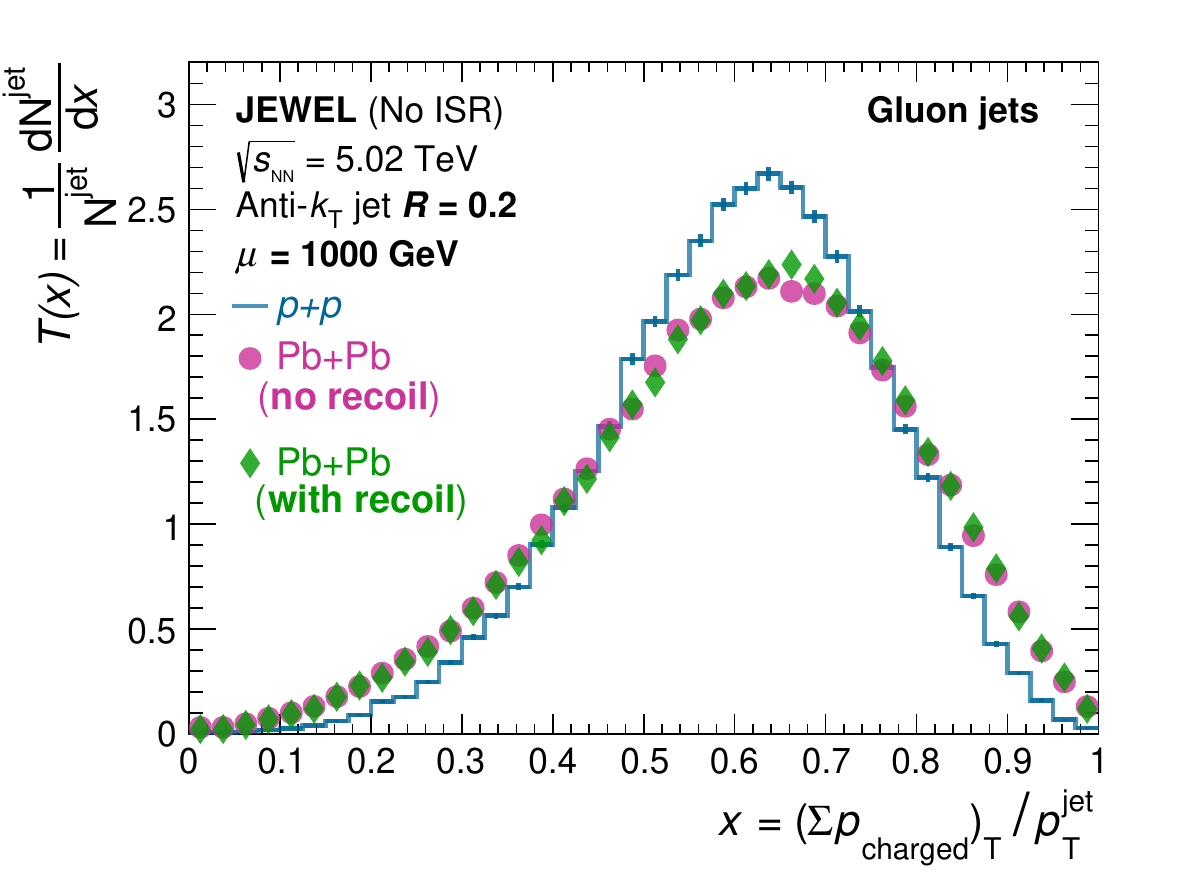}
    \includegraphics[width=0.48\textwidth]{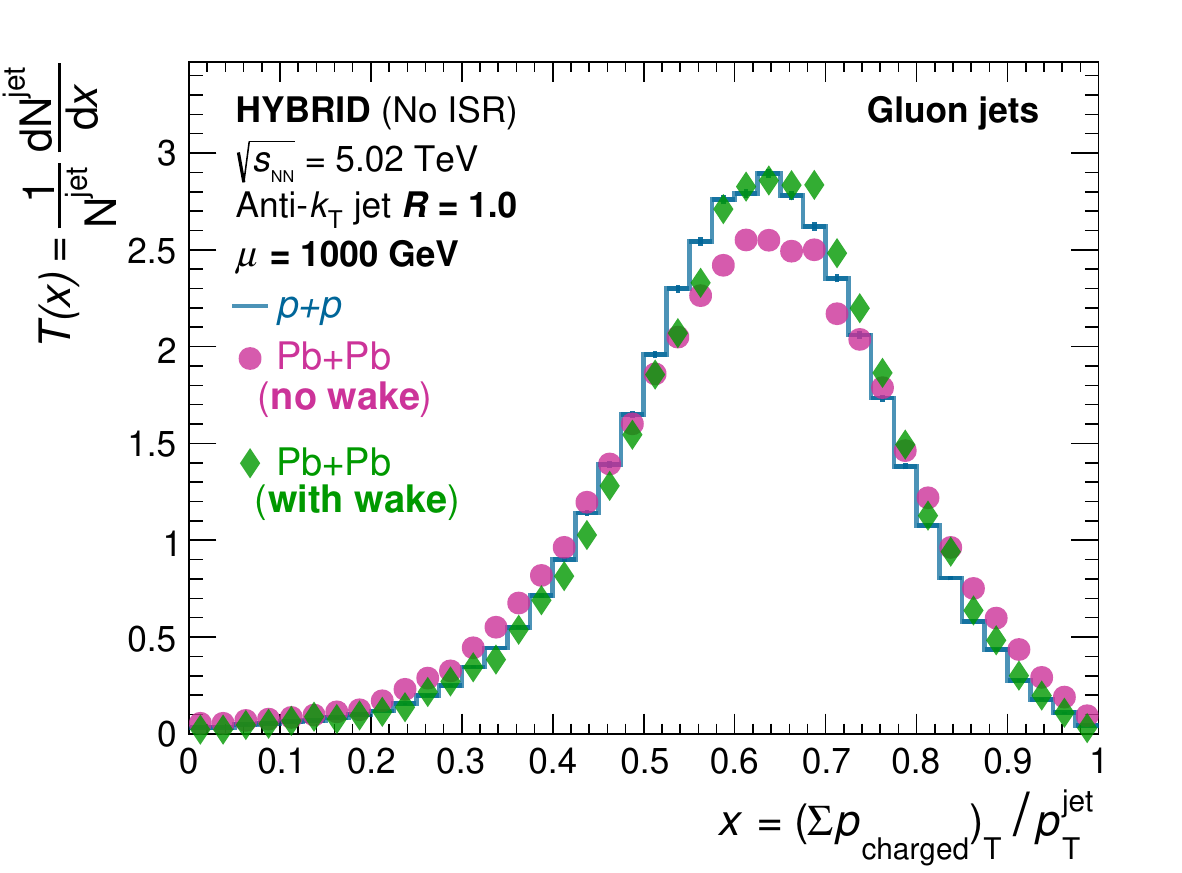}
    \includegraphics[width=0.48\textwidth]{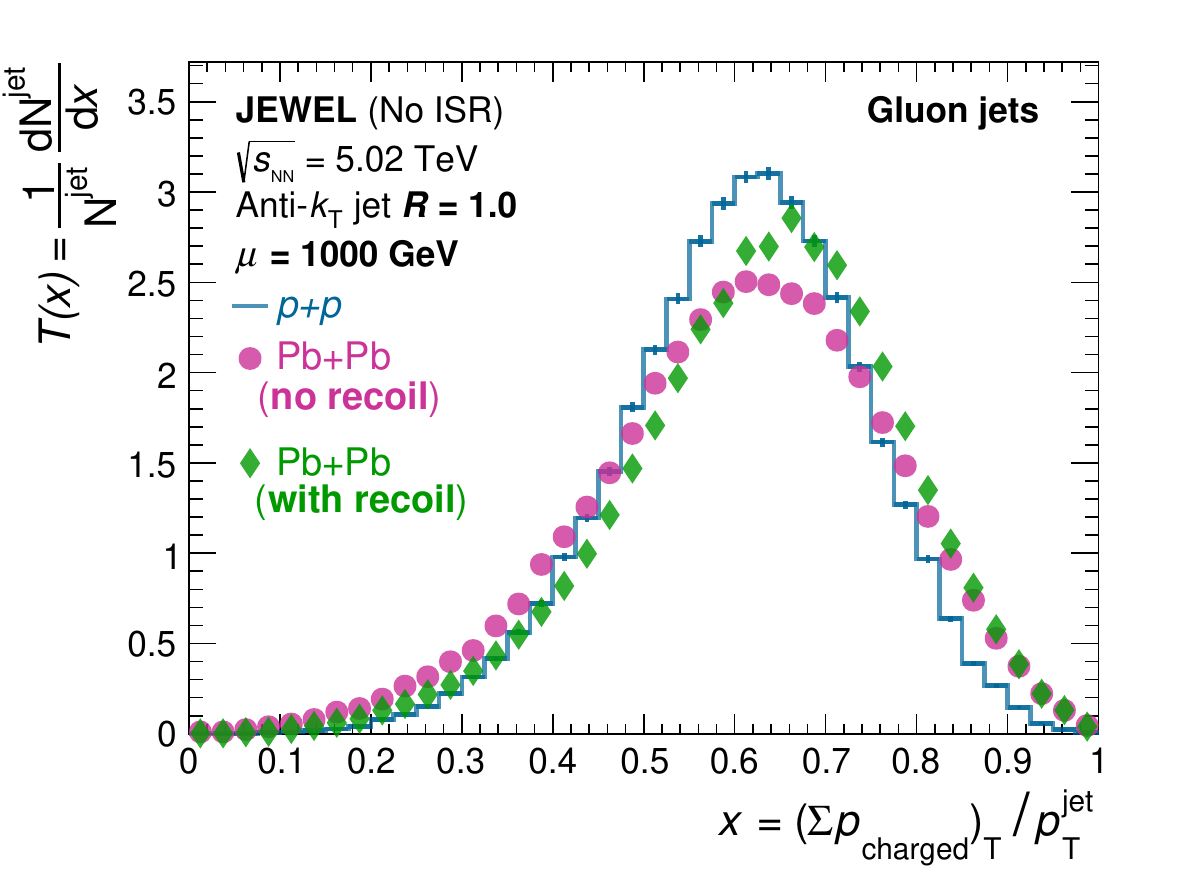}
    \caption{
    Track functions for gluon-initiated jets showing the impact of medium response at $\mu=1000~\mathrm{GeV}$. Results are shown for \hybrid (left panels) and \jewel (right panels), comparing the \pp baseline (blue), \pbpb without medium response (pink circles), and \pbpb with medium response (green diamonds).
    Medium response corresponds to wake contributions in \hybrid and recoil contributions in \jewel. The top and bottom panels show $R=0.2$ and $R=1.0$ jets, respectively.}
\label{fig:jetTF_hybrid_jewel_pthat1000_wakeComp}
\end{figure}

Figures~\ref{fig:jetTF_hybrid_jewel_pthat200_wakeComp} and~\ref{fig:jetTF_hybrid_jewel_pthat1000_wakeComp} show the impact of medium response. For narrow jets with $R=0.2$ at $\mu=200$ GeV, shown in Fig.~\ref{fig:jetTF_hybrid_jewel_pthat200_wakeComp}, the inclusion of wake (\hybrid) or recoil (\jewel) contributions leads to only modest changes in the track function. This indicates that most of the medium response contribution lies at relatively large angles and therefore has limited overlap with a narrow jet cone. In contrast, for $R=1.0$ jets at the same scale $\mu$, medium response produces a very significant modification of the charged fraction distribution. In particular, medium response contributions enhance the distribution at larger charged fractions, modifying the location of the peak. As many other substructure observables, the extraction of track functions can be sensitive to medium response when a sufficiently large jet radius is used. The strong sensitivity of large-radius jets to medium response is another reason why for the goals of this paper we have chosen to use a small jet cone of $R=0.2$. 
At the larger scale $\mu=1000~\mathrm{GeV}$, shown in Fig.~\ref{fig:jetTF_hybrid_jewel_pthat1000_wakeComp}, the \pp and \pbpb distributions without medium response are already close to one another, as discussed in the main text. The scale-dependent reduction of the medium modification is more pronounced in \hybrid than in \jewel, consistent with the moment-level trends shown in Figure~\ref{fig:jetTF_moment_vs_pthat_hybrid_jewel_R2}. Consequently, the additional effects of wake and recoil contributions are also reduced at this scale. The overall impact of medium response is smaller than at $\mu=200~\mathrm{GeV}$, consistent with the expectation that medium-induced modifications are suppressed at larger hard scales.

\clearpage
\section{Additional Studies on Track Function Moments}
\begin{figure}[ht!]
\centering    
    \includegraphics[width=0.48\textwidth]{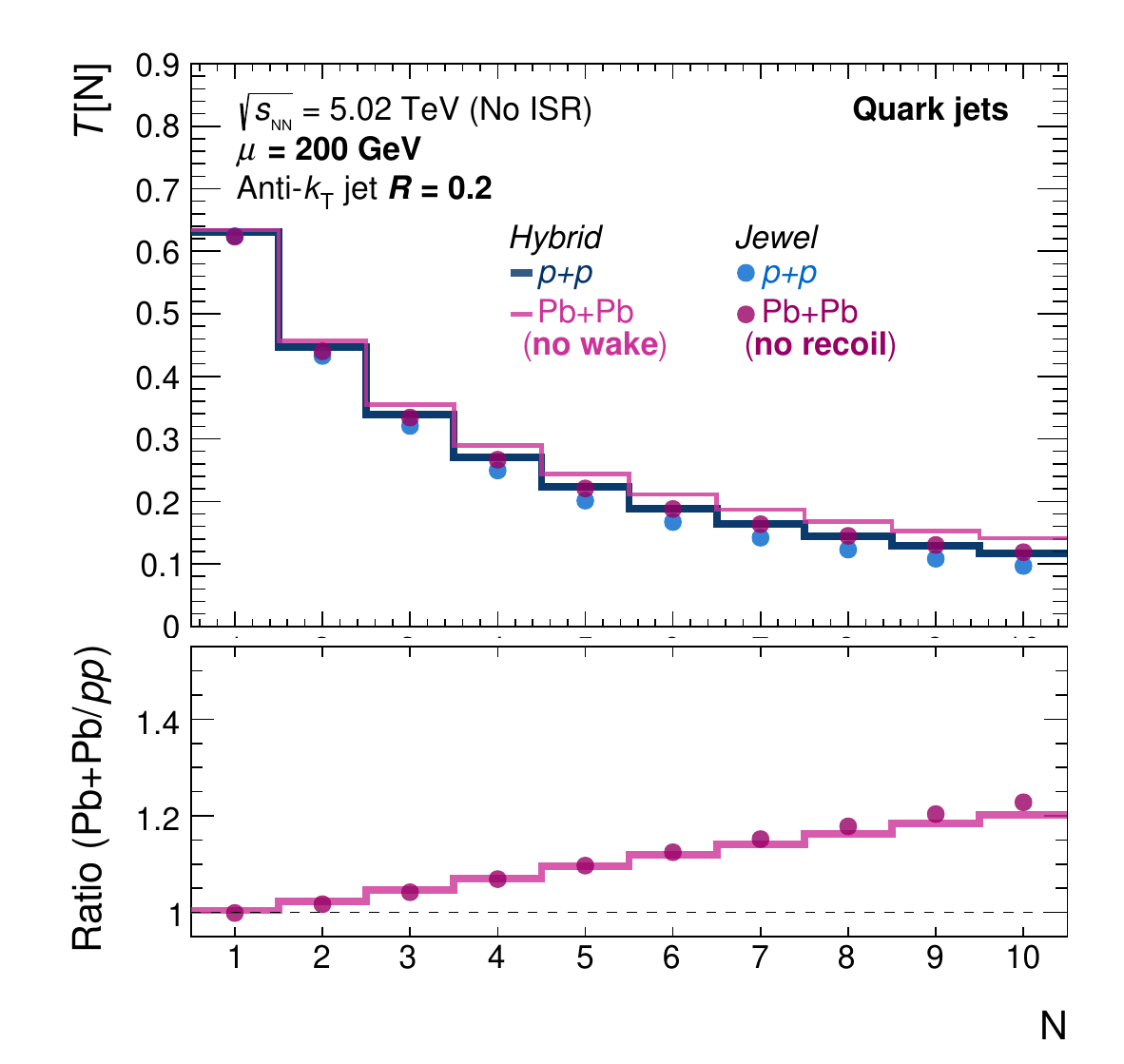}
    \includegraphics[width=0.48\textwidth]{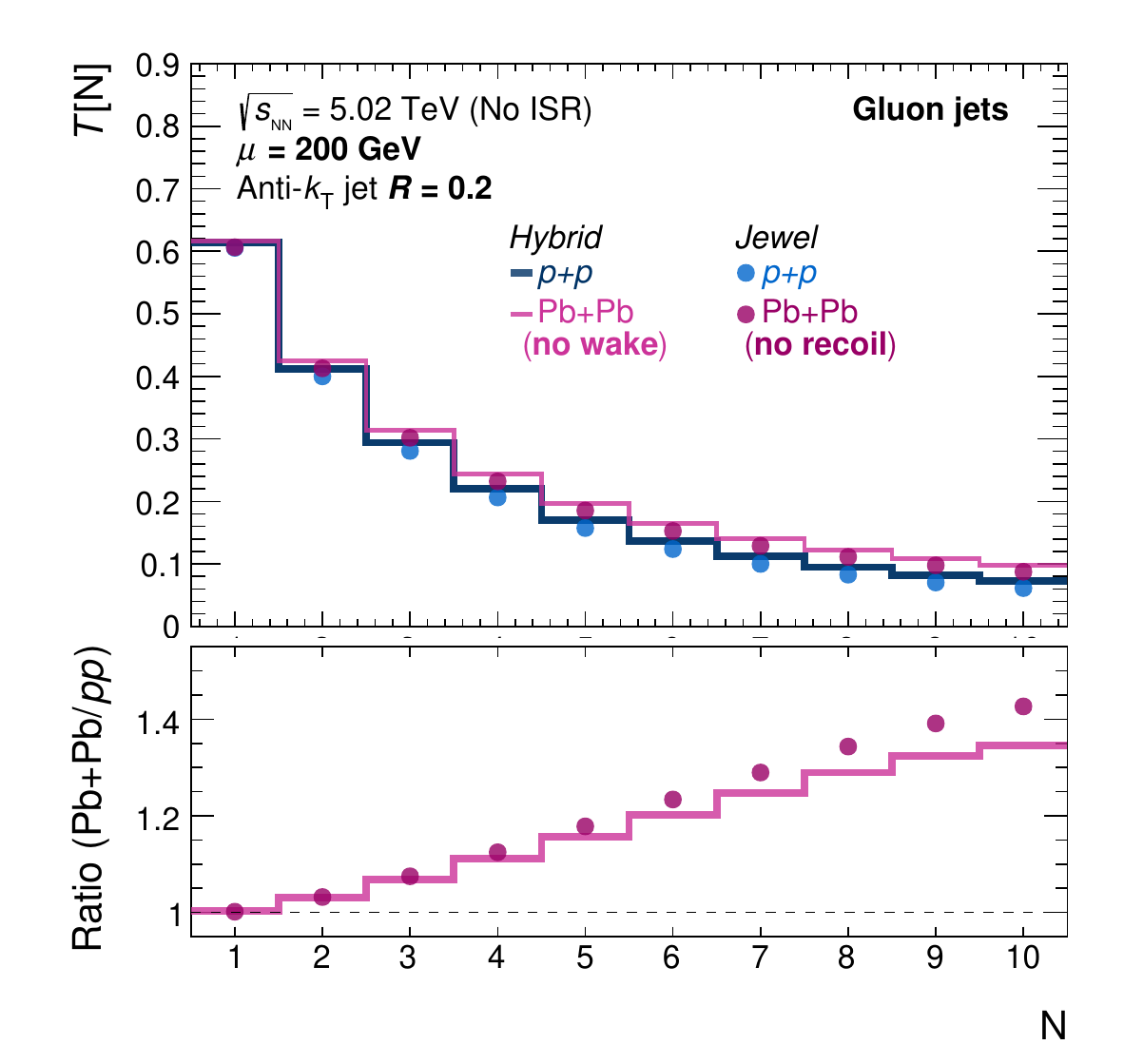}
    \caption{The N-th order moments of the track function for quark-initiated (left) and gluon-initiated (right) jets with R=0.2 at $\mu$ scale of 200 GeV. Results are shown for \hybrid (lines) and \jewel (circles) in \pp (blue) and \pbpb (pink) collisions without the medium response. The bottom panels represent the ratio of \pbpb to \pp.}
    \label{fig:jetTF_moment_vs_N_mu200_R2}
\end{figure}

The same trends are visible in the moments of the track functions. Figure~\ref{fig:jetTF_moment_vs_N_mu200_R2} shows the $N$-th order moments for $R=0.2$ jets at $\mu=200$ GeV. The PbPb to pp ratios increase with the moment order $N$, demonstrating that the medium modification is amplified in the higher moments, as discussed in the main text. Since higher moments give larger weight to the high-$x$ part of the distribution, this behavior reflects a medium-induced reshaping of the charged fraction distribution rather than a simple overall shift. The enhancement is larger for gluon-initiated jets than for quark-initiated jets and is generally stronger in \jewel than in \hybrid. However, at this lower scale $\mu$=200 GeV, \hybrid modifications are still comparable to \jewel's, whereas larger differences between the two models are observed at $\mu=1000~\mathrm{GeV}$, as shown in Fig.~\ref{fig:jetTF_hybrid_jewel_pthat200_R2}.

\begin{figure}[hbtp!]
\centering    
    \includegraphics[width=0.48\textwidth]{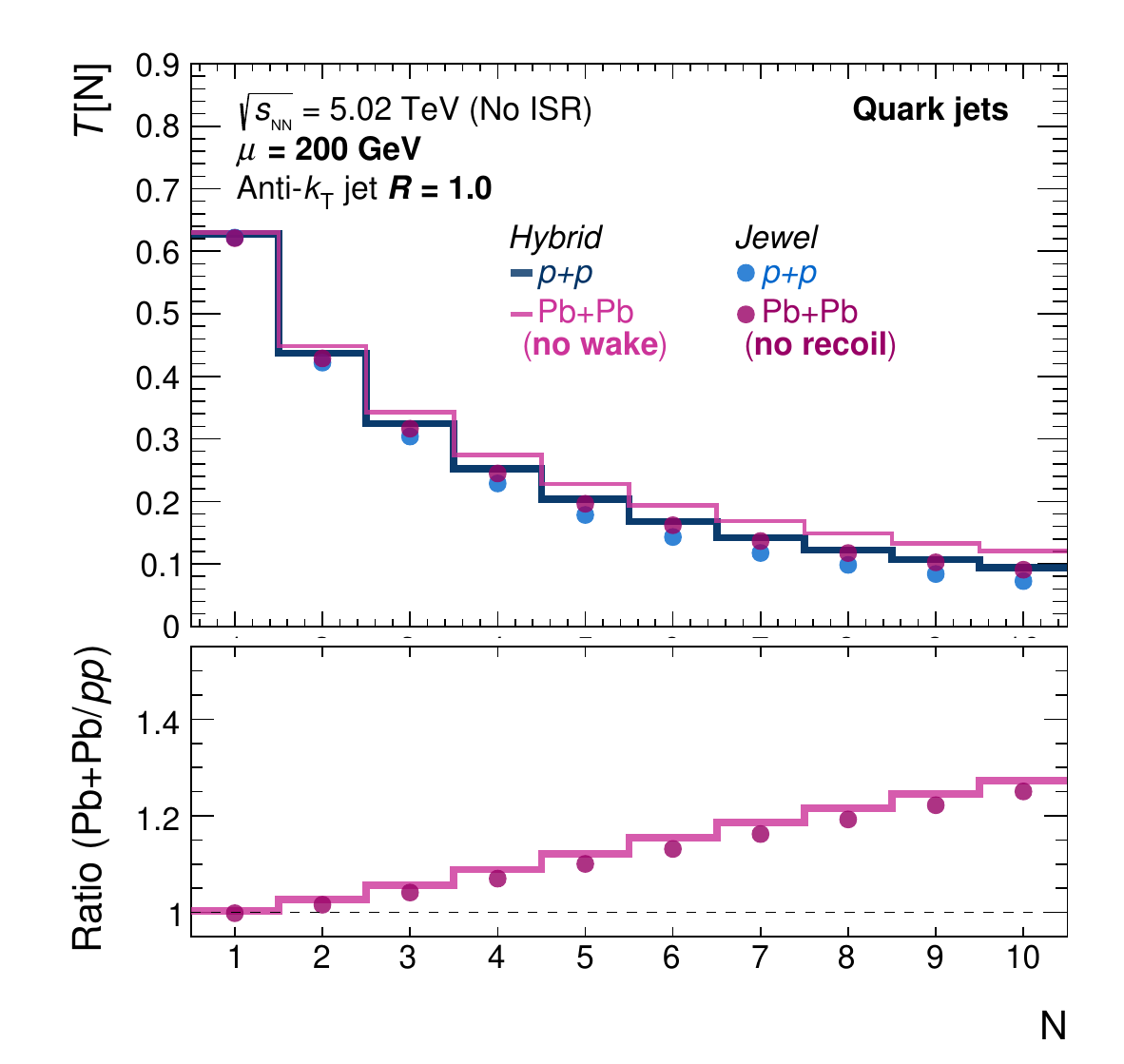}
    \includegraphics[width=0.48\textwidth]{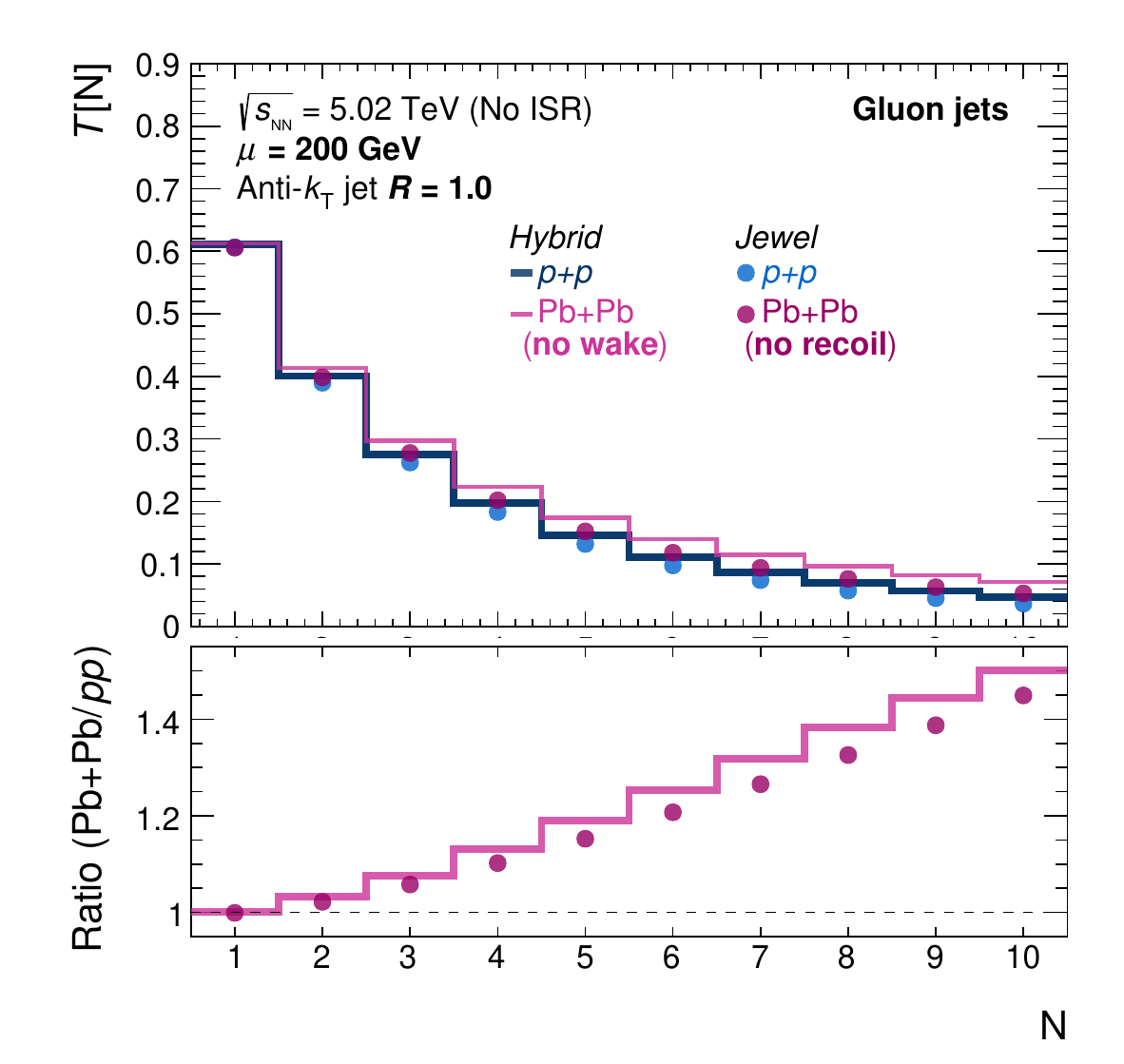}
    \caption{The N-th order moments of the track function for quark-initiated (left) and gluon-initiated (right) jets with R=1.0 at $\mu$ scale of 200 GeV. Results are shown for \hybrid (lines) and \jewel (circles) in \pp (blue) and \pbpb (pink) collisions without the medium response. The bottom panels represent the ratio of \pbpb to \pp.}
    \label{fig:jetTF_moment_vs_N_mu200_R10}
\end{figure}

\begin{figure}[hbtp!]
\centering    
    \includegraphics[width=0.48\textwidth]{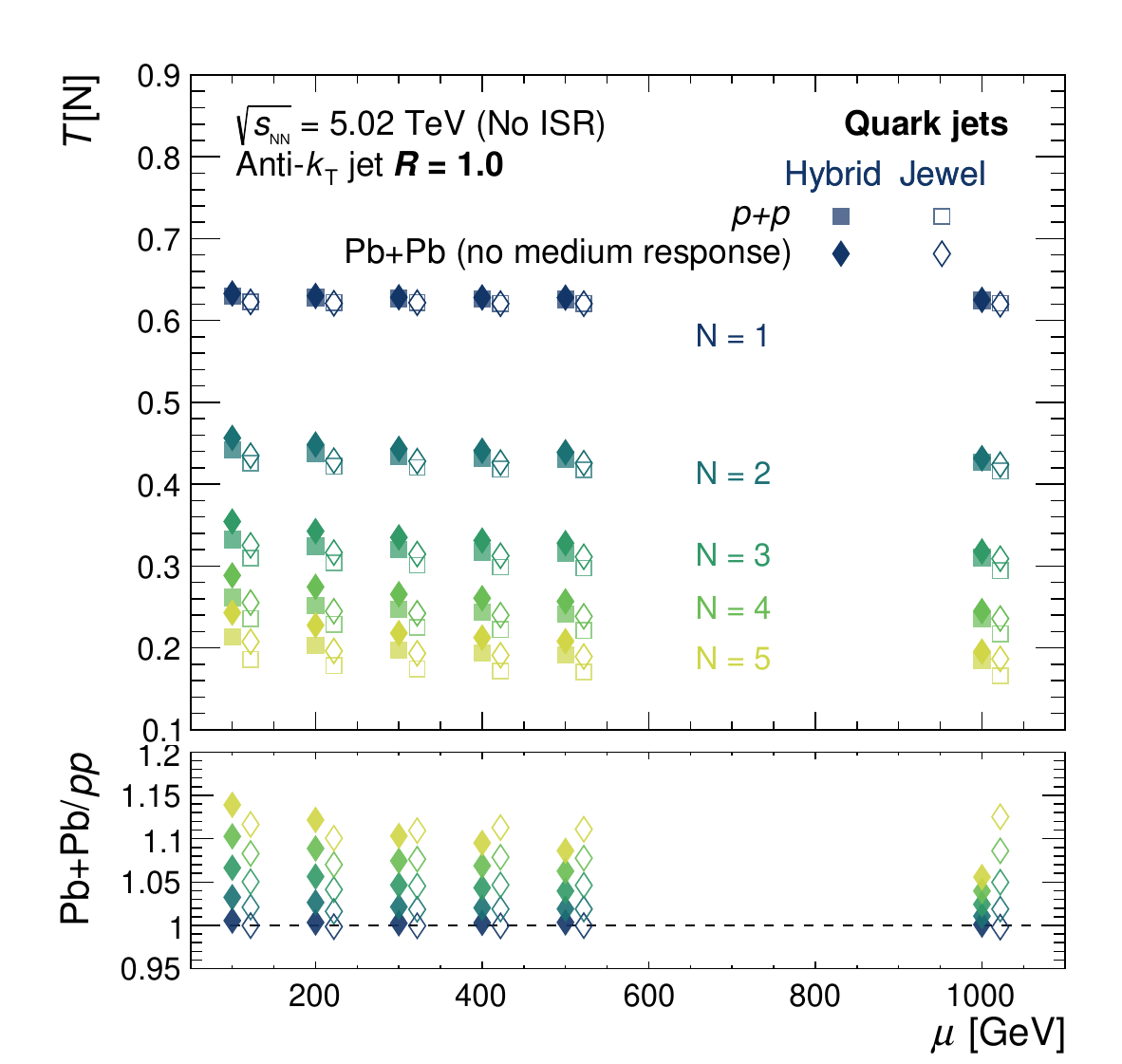}
    \includegraphics[width=0.48\textwidth]{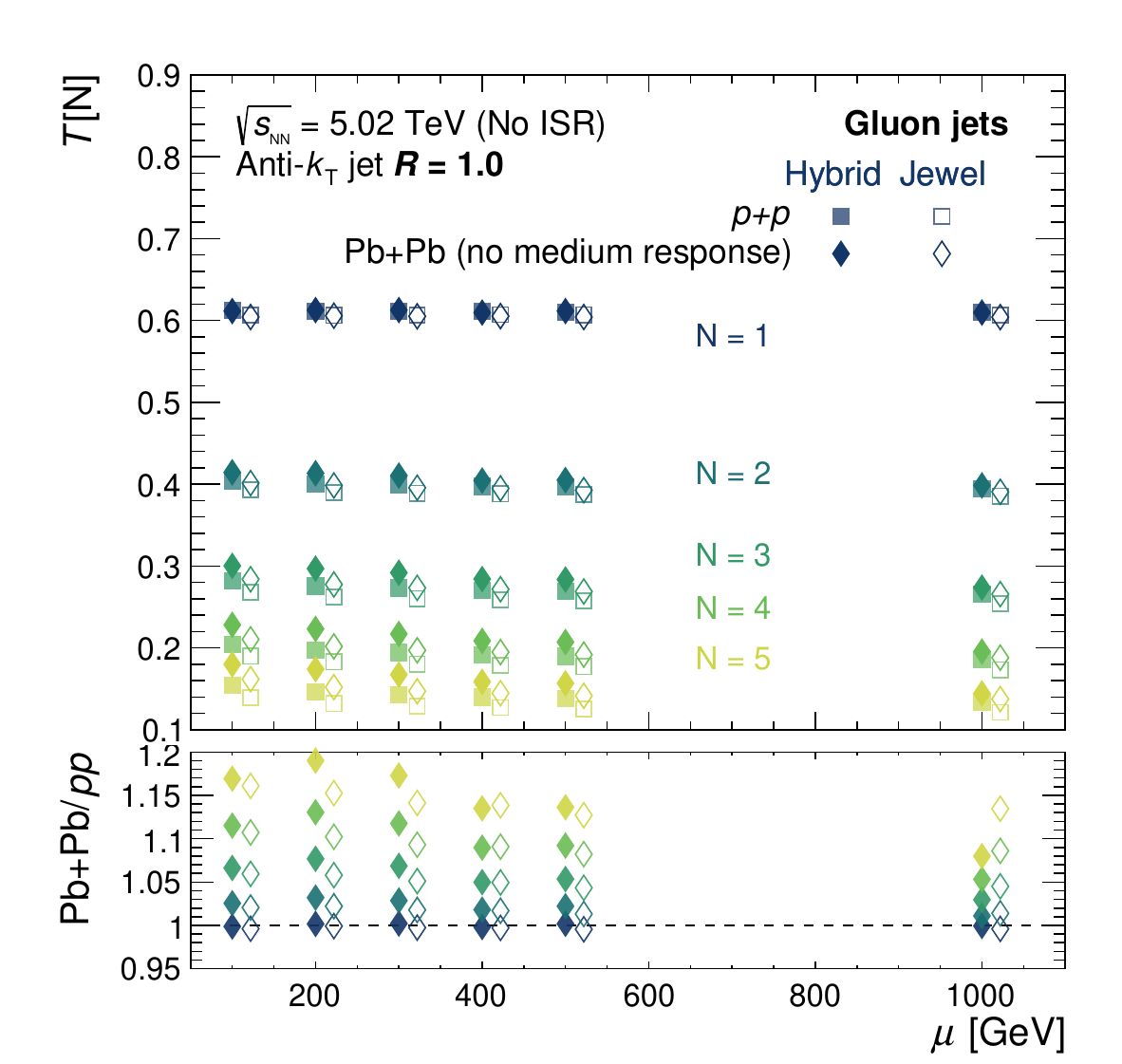}
    \caption{The N-th order moments of the track function as a function of $\mu$ for \pp collisions and \pbpb without wake in \hybrid (blue) and \jewel (pink) for $R$=1.0 jets. Left panel is for quark-initiated jets and right panel for gluon-initiated jets.
    Different colors represent different order of moments. 
    The bottom panels represent the ratio of \pbpb to \pp.}
    \label{fig:jetTF_moment_vs_pthat_hybrid_jewel_R10}
\end{figure}

Figures~\ref{fig:jetTF_moment_vs_N_mu200_R10} and~\ref{fig:jetTF_moment_vs_pthat_hybrid_jewel_R10} extend this study to $R=1.0$ jets. Figure~\ref{fig:jetTF_moment_vs_N_mu200_R10} shows the $N$th-order moments at the same scale, $\mu=200~\mathrm{GeV}$. The higher moments still show \pbpb to \pp ratios above unity, with a larger enhancement at higher $N$. Comparing the two jet radii between Fig.~\ref{fig:jetTF_moment_vs_N_mu200_R2} and ~\ref{fig:jetTF_moment_vs_N_mu200_R10}, the $N$ dependence in \jewel is qualitatively similar between $R=0.2$ and $R=1.0$. In contrast, \hybrid shows a slightly more pronounced increase of the \pbpb to \pp ratio for $R=1.0$, indicating a stronger jet-radius dependence of the medium modification in \hybrid. 

The scale dependence in Figure~\ref{fig:jetTF_moment_vs_pthat_hybrid_jewel_R10} shows that this qualitative behavior of larger \pbpb to \pp ratios above unity, with a larger enhancement at higher $N$ persists over the full range of $\mu$ studied. The moment ratios remain close to unity for the first moment but increase for higher moments, indicating that the mean charged fraction is relatively stable while the shape and fluctuations of the distribution are more sensitive to the medium. Compared to the smaller $R$ values used in the main text, one can observe that medium modifications for both models are larger for these wider $R=1$ jets. This is consistent with the expectation that wider jets possess more sources of energy loss (provided that they are resolved by the medium) and are thus more quenched as a whole.


\clearpage
\section{Scale Evolution of Higher Order Cumulants}
\begin{figure}[hbtp!]
    \centering
       \includegraphics[width=0.48\linewidth]{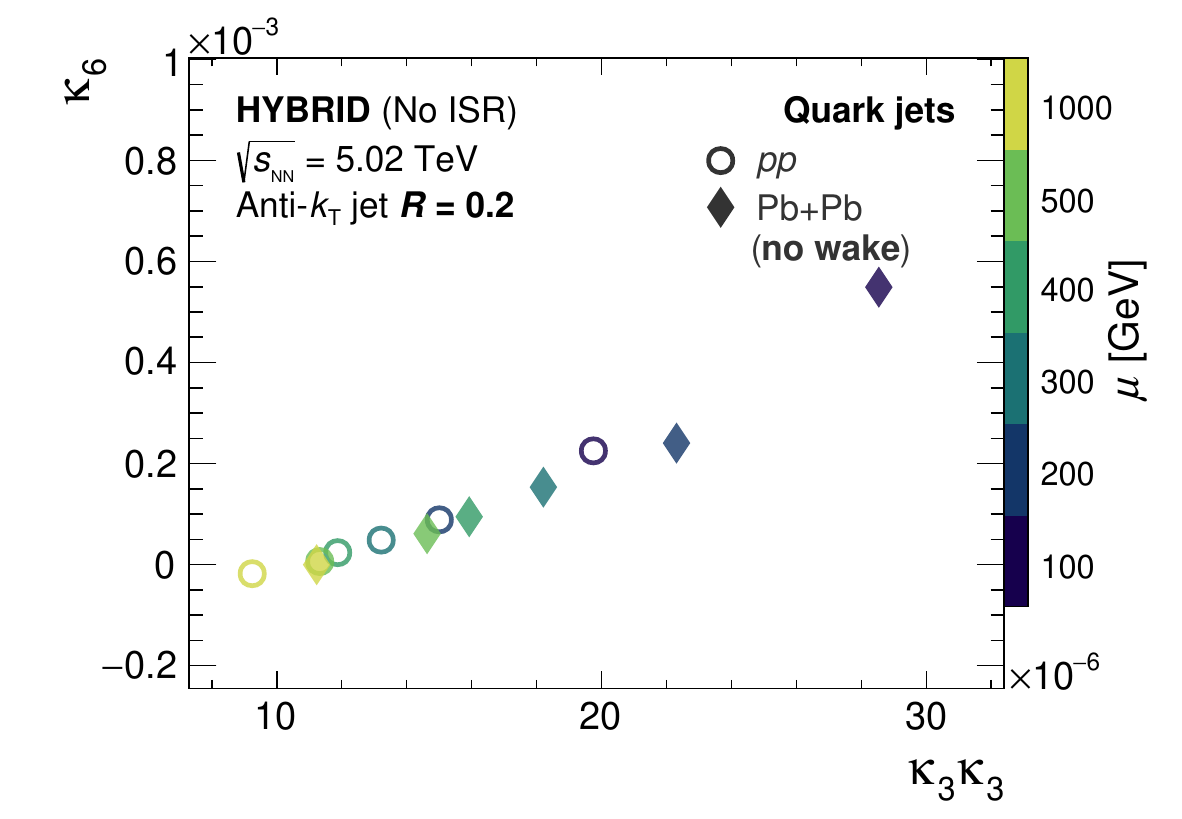}
    \includegraphics[width=0.48\linewidth]{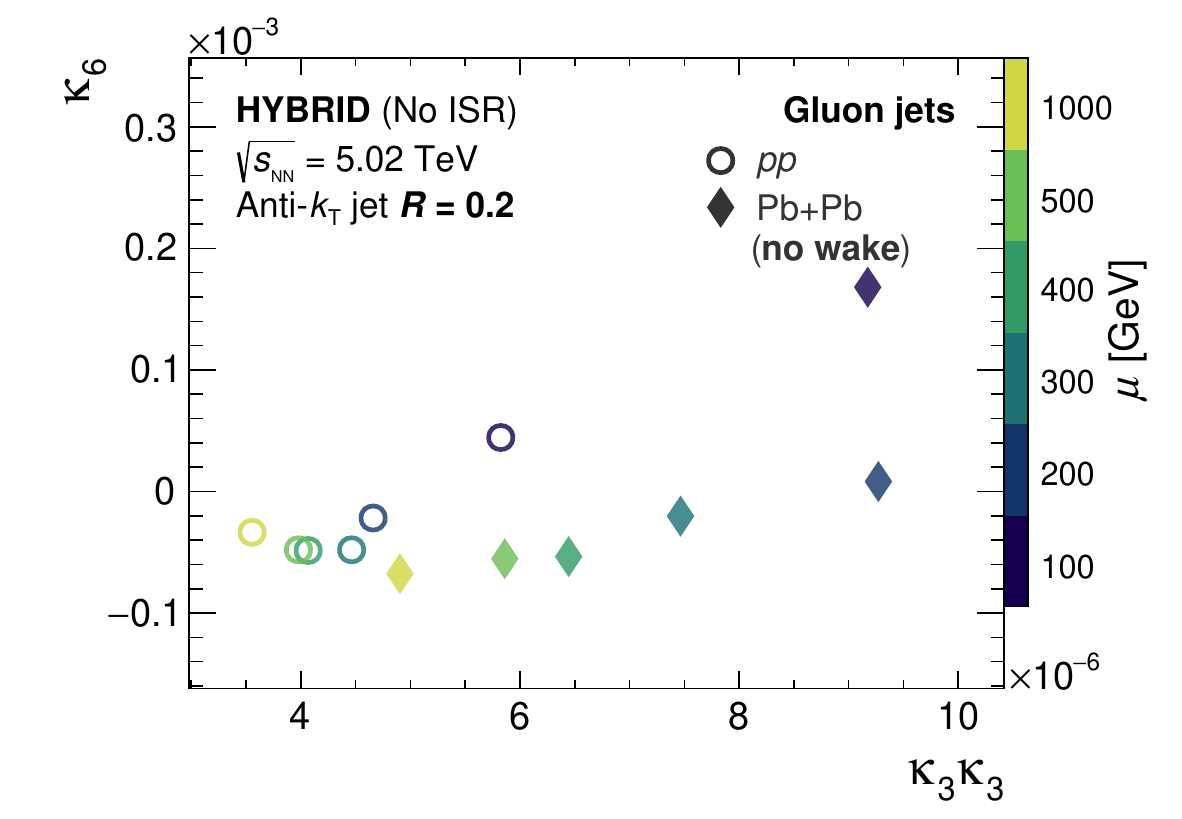}
    \includegraphics[width=0.48\linewidth]{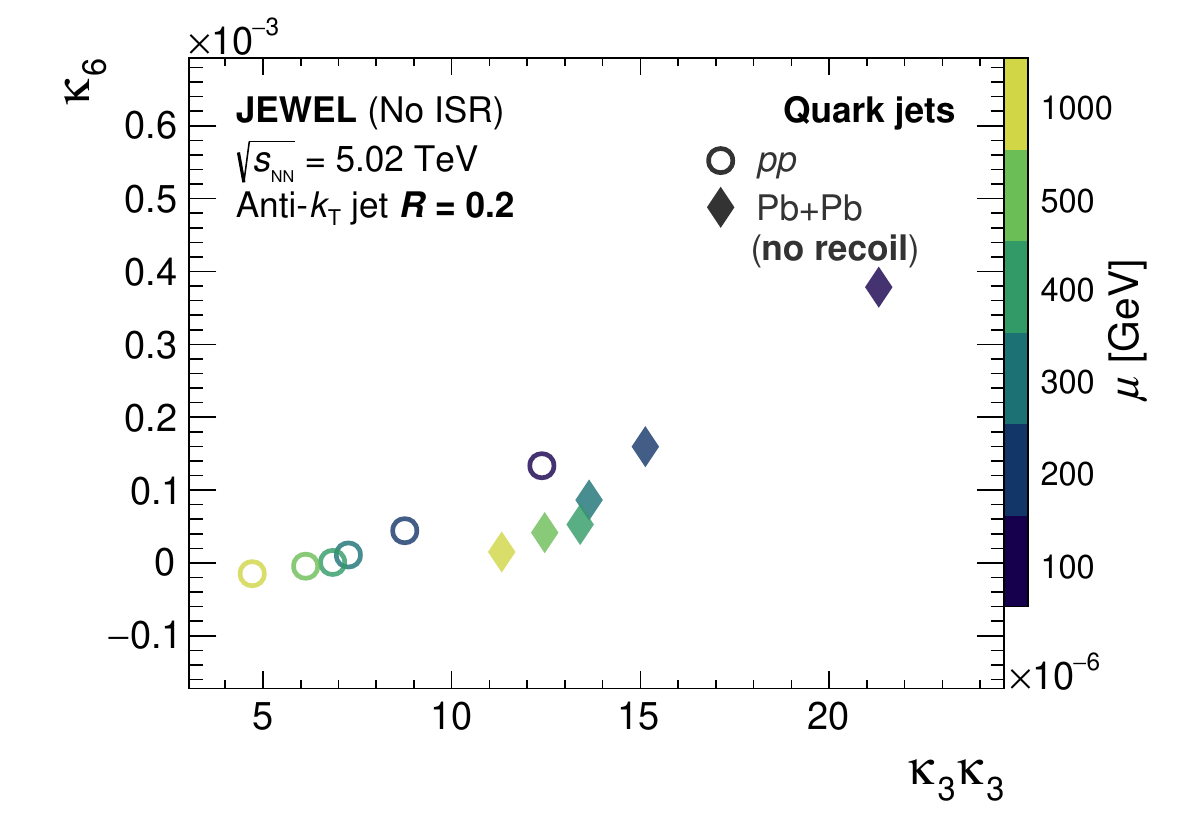}
    \includegraphics[width=0.48\linewidth]{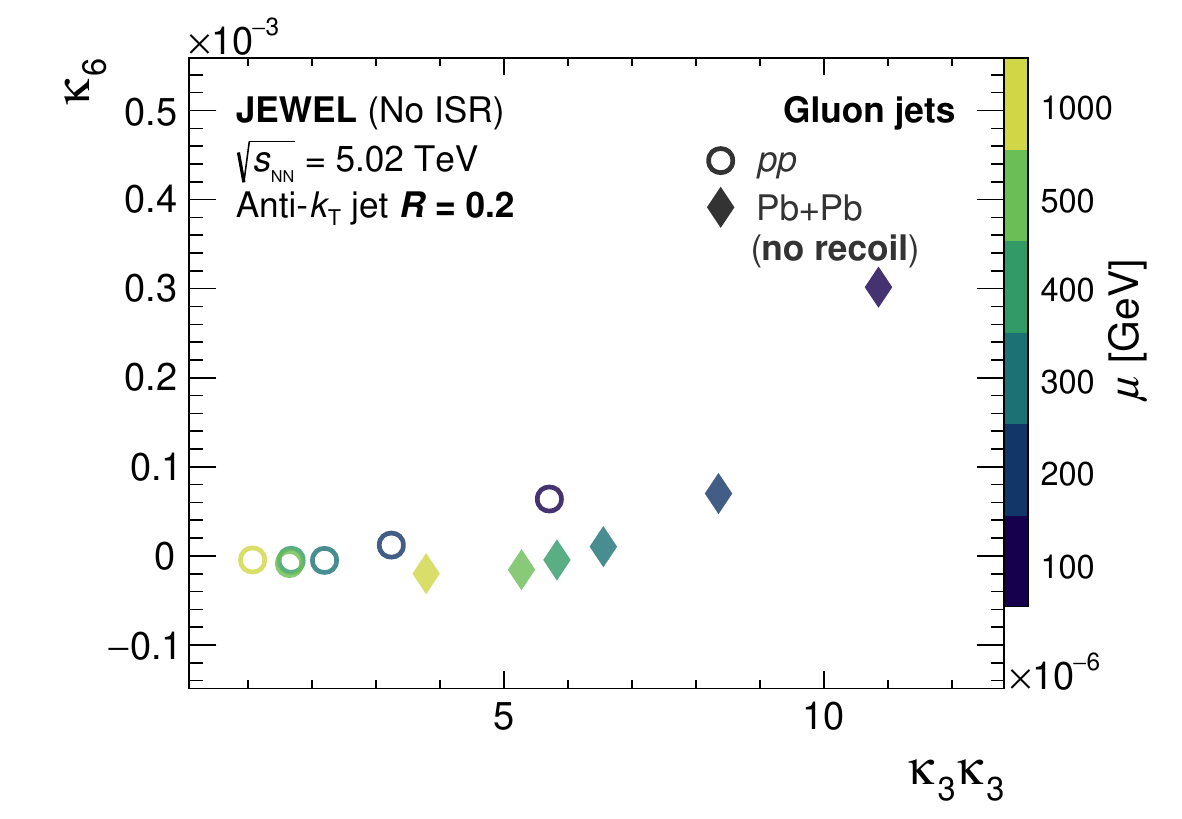}

    \caption{ $\kappa_6$ vs $\kappa_3^2$ cumulants for \pp (open circle) and \pbpb (filled diamond). Results from the \hybrid (top) and \jewel (bottom) are shown for quark-initiated jets (left) and gluon-initiated jets (right) for $R$=0.2 jets. }
    \label{fig:jet_tf_k6_vs_k32_R2}
\end{figure}

\begin{figure}[hbtp!]
    \centering
       \includegraphics[width=0.48\linewidth]{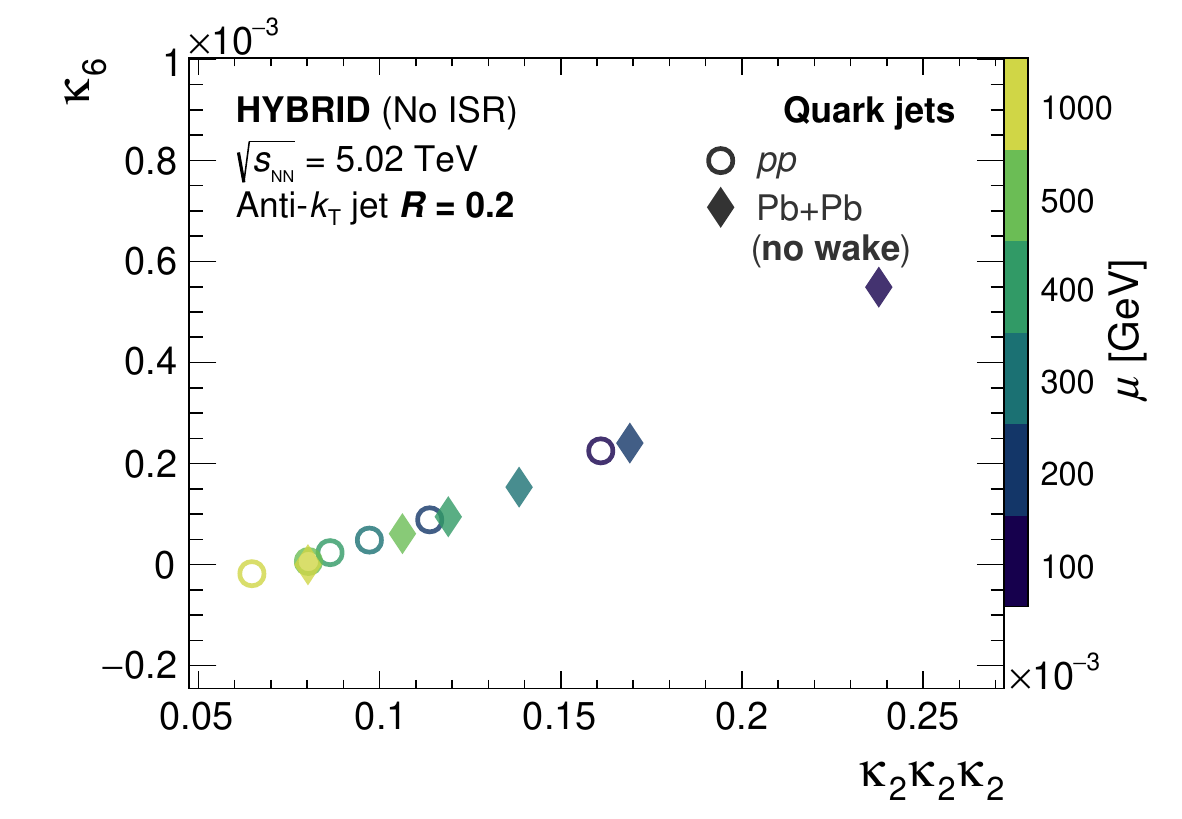}
    \includegraphics[width=0.48\linewidth]{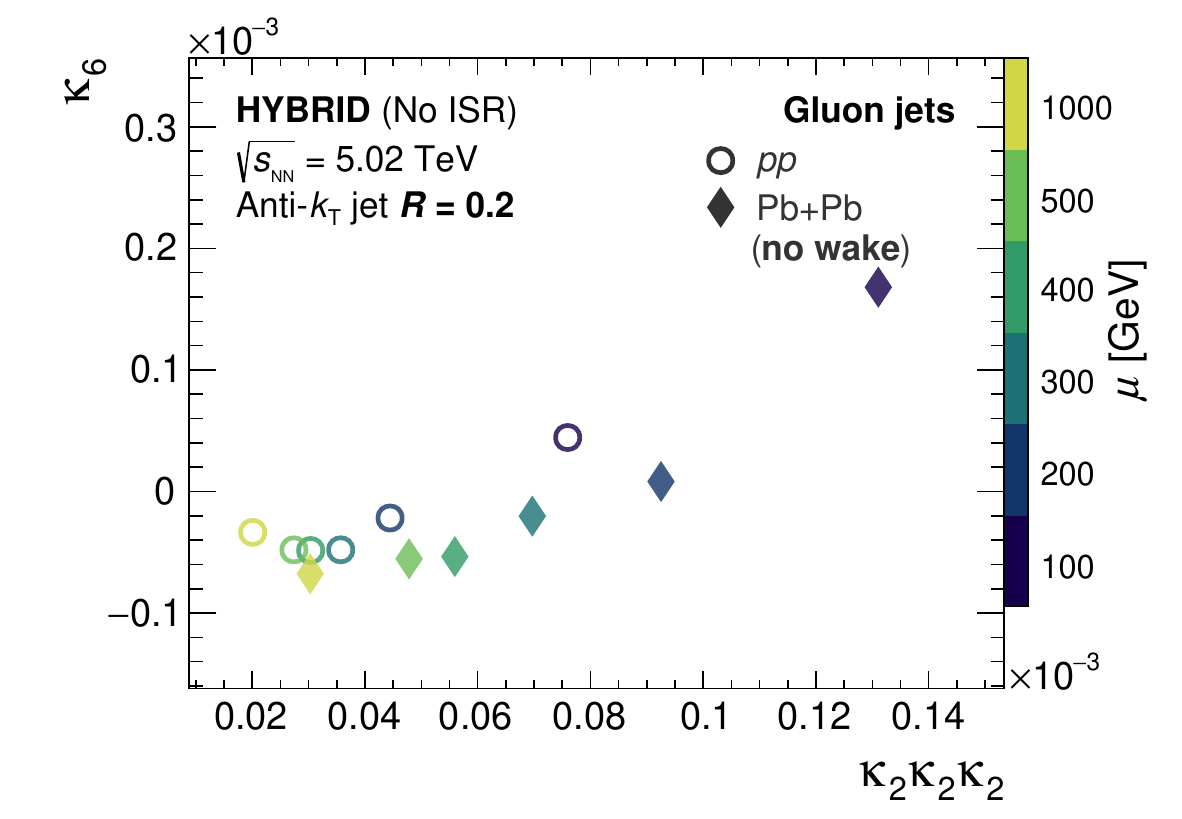}
    \includegraphics[width=0.48\linewidth]{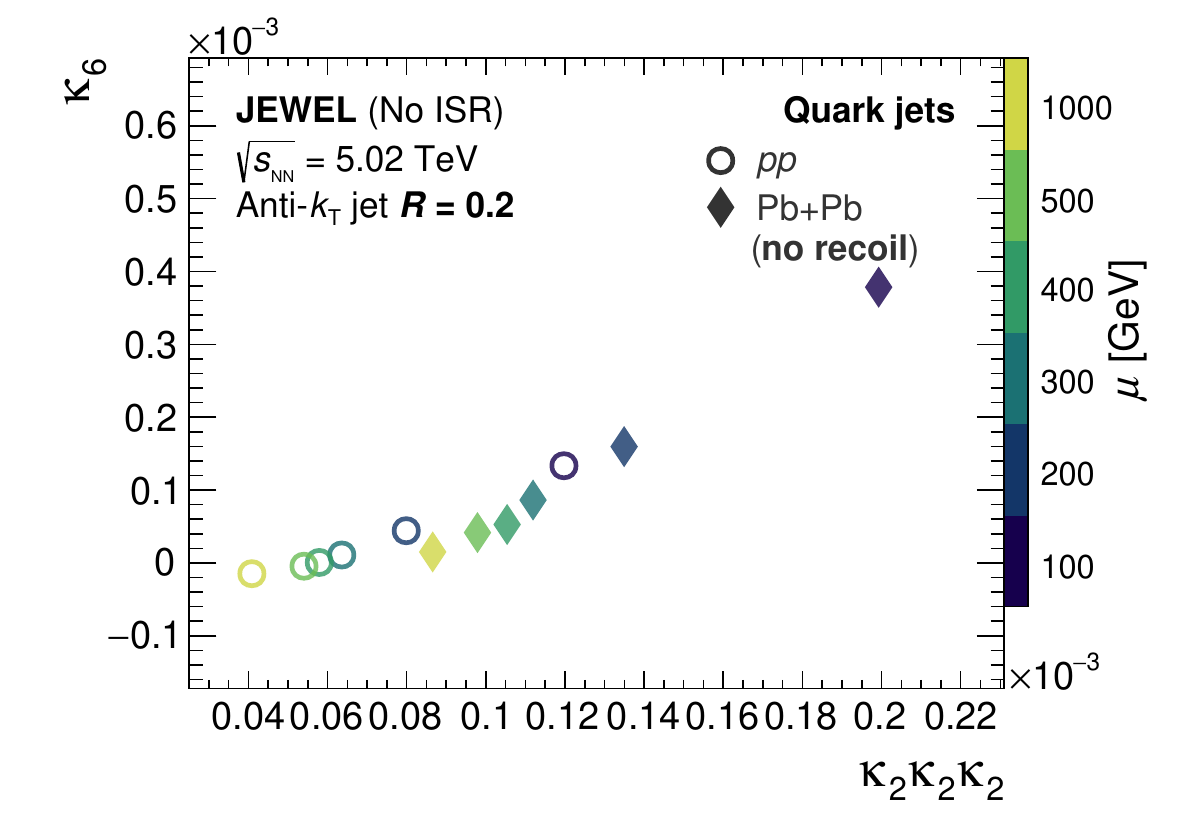}
    \includegraphics[width=0.48\linewidth]{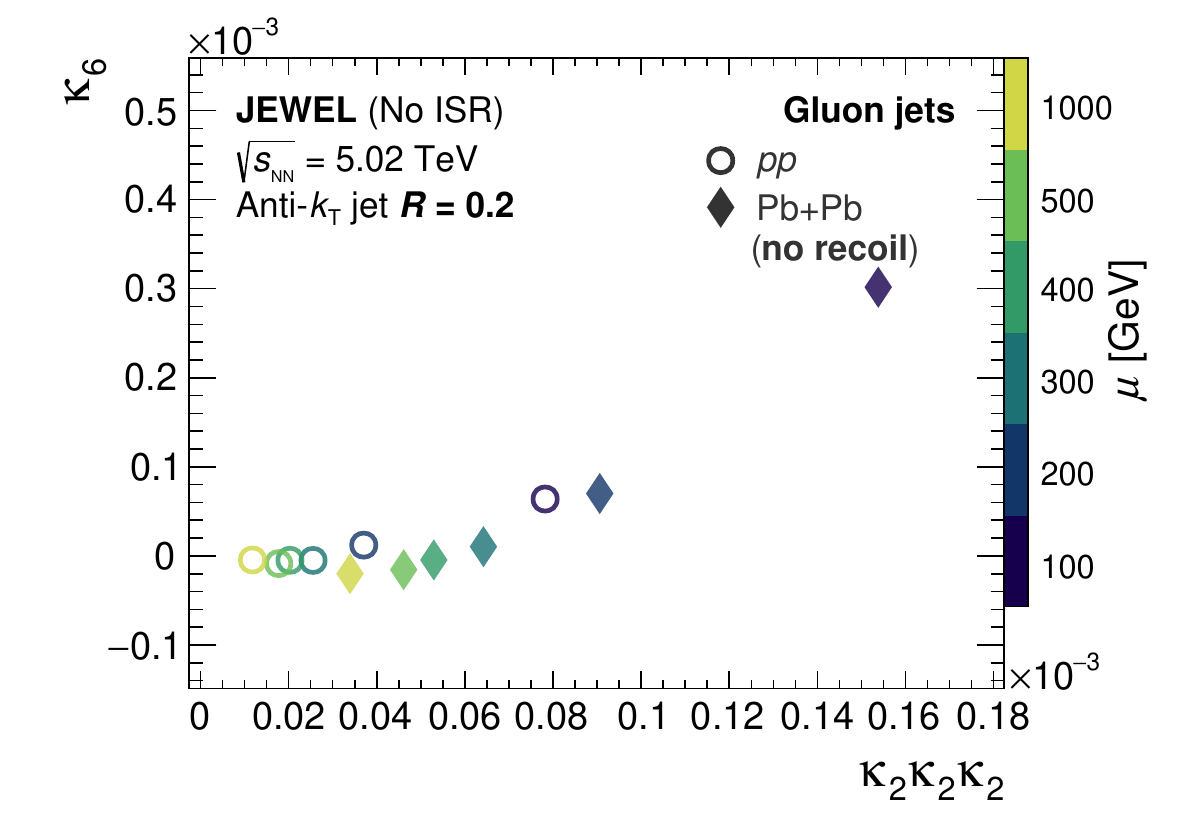}

    \caption{ $\kappa_6$ vs $\kappa_2^3$ cumulants for \pp (open circle) and \pbpb (filled diamond). Results from the \hybrid (top) and \jewel (bottom) are shown for quark-initiated jets (left) and gluon-initiated jets (right) for $R$=0.2 jets.}
    \label{fig:jet_tf_k6_vs_k222_R2}
\end{figure}

Finally, Figures~\ref{fig:jet_tf_k6_vs_k32_R2} and~\ref{fig:jet_tf_k6_vs_k222_R2} show the evolution of the sixth cumulant  in relation to lower cumulant combinations, $\kappa_3^2$ and $\kappa_2^3$, for $R=0.2$ jets. The pp and PbPb points follow broadly similar trajectories as a function of $\mu$, supporting the conclusion that the qualitative structure of the cumulant flow remains close to the vacuum case. However, the sixth cumulant exhibits larger scatter and larger deviations between pp and PbPb than the lower cumulants shown in the main text. Since high-order cumulants are more sensitive to the tails of the track function and to statistical fluctuations, these results should be interpreted as qualitative. A more precise assessment of the highest cumulants will require larger event samples and a dedicated study of statistical uncertainties.

